\documentclass[conference,a4paper]{IEEEtran}
\IEEEoverridecommandlockouts
\usepackage[acronyms,nonumberlist,nopostdot,nomain,nogroupskip,acronymlists={hidden}]{glossaries}
\newglossary[algh]{hidden}{acrh}{acnh}{Hidden Acronyms}

\usepackage{cite}
\usepackage{amsmath,amssymb,amsfonts}
\usepackage{algorithmic}
\usepackage{textcomp}
\usepackage{xcolor}

\usepackage{booktabs}
\usepackage{makecell}

\usepackage{tikz}
\usepackage{pgfplots}
\pgfplotsset{compat=newest}
\pgfplotsset{plot coordinates/math parser=false}
\newlength\fheight
\newlength\fwidth
\usetikzlibrary{plotmarks,patterns,decorations.pathreplacing,backgrounds,calc,arrows,arrows.meta,spy,matrix,scopes}
\usepgfplotslibrary{patchplots,groupplots}
\usepackage{tikzscale}
\pgfplotsset{compat=1.18}

\def\BibTeX{{\rm B\kern-.05em{\sc i\kern-.025em b}\kern-.08em
    T\kern-.1667em\lower.7ex\hbox{E}\kern-.125emX}}

\expandafter\def\csname ver@fixltx2e.sty\endcsname{}

\newif\ifexttikz
\exttikzfalse

\ifexttikz
	\usetikzlibrary{external}
	\usepackage{fontspec}
\fi

\usepackage[font=footnotesize]{caption}
\usepackage{subcaption}
\usepackage{soul}

\newacronym{3gpp}{3GPP}{3rd Generation Partnership Project}
\newacronym{4g}{4G}{4th generation}
\newacronym{5g}{5G}{5th generation}
\newacronym{6g}{6G}{6th generation}
\newacronym{5gc}{5GC}{5G Core}
\newacronym{adc}{ADC}{Analog to Digital Converter}
\newacronym{aerpaw}{AERPAW}{Aerial Experimentation and Research Platform for Advanced Wireless}
\newacronym{ai}{AI}{Artificial Intelligence}
\newacronym{aimd}{AIMD}{Additive Increase Multiplicative Decrease}
\newacronym{am}{AM}{Acknowledged Mode}
\newacronym{amc}{AMC}{Adaptive Modulation and Coding}
\newacronym{amf}{AMF}{Access and Mobility Management Function}
\newacronym{aops}{AOPS}{Adaptive Order Prediction Scheduling}
\newacronym{api}{API}{Application Programming Interface}
\newacronym{apn}{APN}{Access Point Name}
\newacronym{ap}{AP}{Application Protocol}
\newacronym{aqm}{AQM}{Active Queue Management}
\newacronym{ausf}{AUSF}{Authentication Server Function}
\newacronym{avc}{AVC}{Advanced Video Coding}
\newacronym{awgn}{AGWN}{Additive White Gaussian Noise}
\newacronym{balia}{BALIA}{Balanced Link Adaptation Algorithm}
\newacronym{bbu}{BBU}{Base Band Unit}
\newacronym{bdp}{BDP}{Bandwidth-Delay Product}
\newacronym{ber}{BER}{Bit Error Rate}
\newacronym{bf}{BF}{Beamforming}
\newacronym{bler}{BLER}{Block Error Rate}
\newacronym{brr}{BRR}{Bayesian Ridge Regressor}
\newacronym{bs}{BS}{Base Station}
\newacronym{bsr}{BSR}{Buffer Status Report}
\newacronym{bss}{BSS}{Business Support System}
\newacronym{ca}{CA}{Carrier Aggregation}
\newacronym{caas}{CaaS}{Connectivity-as-a-Service}
\newacronym{cb}{CB}{Code Block}
\newacronym{cc}{CC}{Congestion Control}
\newacronym{ccid}{CCID}{Congestion Control ID}
\newacronym{cco}{CC}{Carrier Component}
\newacronym{cdd}{CDD}{Cyclic Delay Diversity}
\newacronym{cdf}{CDF}{Cumulative Distribution Function}
\newacronym{cdn}{CDN}{Content Distribution Network}
\newacronym{cir}{CIR}{Channel Impulse Response}
\newacronym{cli}{CLI}{Command-line Interface}
\newacronym{cn}{CN}{Core Network}
\newacronym{cnn}{CNN}{Convolutional Neural Network}
\newacronym{codel}{CoDel}{Controlled Delay Management}
\newacronym{comac}{COMAC}{Converged Multi-Access and Core}
\newacronym{cord}{CORD}{Central Office Re-architected as a Datacenter}
\newacronym{cornet}{CORNET}{COgnitive Radio NETwork}
\newacronym{cosmos}{COSMOS}{Cloud Enhanced Open Software Defined Mobile Wireless Testbed for City-Scale Deployment}
\newacronym{cots}{COTS}{Commercial Off-the-Shelf}
\newacronym{cp}{CP}{Control Plane}
\newacronym{cyp}{CP}{Cyclic Prefix}
\newacronym{up}{UP}{User Plane}
\newacronym{cpu}{CPU}{Central Processing Unit}
\newacronym{cqi}{CQI}{Channel Quality Information}
\newacronym{cql}{CQL}{Conservative Q-Learning}
\newacronym{cr}{CR}{Cognitive Radio}
\newacronym{cran}{CRAN}{Cloud \gls{ran}}
\newacronym{crs}{CRS}{Cell Reference Signal}
\newacronym{csi}{CSI}{Channel State Information}
\newacronym{csirs}{CSI-RS}{Channel State Information - Reference Signal}
\newacronym{cu}{CU}{Central Unit}
\newacronym{cucp}{CU-CP}{Central Unit Control Plane}
\newacronym{cuup}{CU-UP}{Central Unit User Plane}
\newacronym{d2tcp}{D$^2$TCP}{Deadline-aware Data center TCP}
\newacronym{d3}{D$^3$}{Deadline-Driven Delivery}
\newacronym{dac}{DAC}{Digital to Analog Converter}
\newacronym{dag}{DAG}{Directed Acyclic Graph}
\newacronym{das}{DAS}{Distributed Antenna System}
\newacronym{dash}{DASH}{Dynamic Adaptive Streaming over HTTP}
\newacronym{dc}{DC}{Dual Connectivity}
\newacronym{dccp}{DCCP}{Datagram Congestion Control Protocol}
\newacronym{dce}{DCE}{Direct Code Execution}
\newacronym{dci}{DCI}{Downlink Control Information}
\newacronym{dctcp}{DCTCP}{Data Center TCP}
\newacronym{dl}{DL}{Downlink}
\newacronym{dmr}{DMR}{Deadline Miss Ratio}
\newacronym{dmrs}{DMRS}{DeModulation Reference Signal}
\newacronym{dqn}{DQN}{Deep Q-Network}
\newacronym{drlcc}{DRL-CC}{Deep Reinforcement Learning Congestion Control}
\newacronym{drs}{DRS}{Discovery Reference Signal}
\newacronym{du}{DU}{Distributed Unit}
\newacronym{ee}{EE}{Energy Efficiency}
\newacronym{e2e}{E2E}{end-to-end}
\newacronym{earfcn}{EARFCN}{E-UTRA Absolute Radio Frequency Channel Number}
\newacronym{ecaas}{ECaaS}{Edge-Cloud-as-a-Service}
\newacronym{ecn}{ECN}{Explicit Congestion Notification}
\newacronym{edf}{EDF}{Earliest Deadline First}
\newacronym{embb}{eMBB}{Enhanced Mobile Broadband}
\newacronym{empower}{EMPOWER}{EMpowering transatlantic PlatfOrms for advanced WirEless Research}
\newacronym{enb}{eNB}{evolved Node Base}
\newacronym{endc}{EN-DC}{E-UTRAN-\gls{nr} \gls{dc}}
\newacronym{epc}{EPC}{Evolved Packet Core}
\newacronym{eps}{EPS}{Evolved Packet System}
\newacronym{es}{ES}{Edge Server}
\newacronym{etsi}{ETSI}{European Telecommunications Standards Institute}
\newacronym[firstplural=Estimated Times of Arrival (ETAs)]{eta}{ETA}{Estimated Time of Arrival}
\newacronym{eutran}{E-UTRAN}{Evolved Universal Terrestrial Access Network}
\newacronym{faas}{FaaS}{Function-as-a-Service}
\newacronym{fapi}{FAPI}{Functional Application Platform Interface}
\newacronym{fdd}{FDD}{Frequency Division Duplexing}
\newacronym{fdm}{FDM}{Frequency Division Multiplexing}
\newacronym{fdma}{FDMA}{Frequency Division Multiple Access}
\newacronym{fed4fire}{FED4FIRE+}{Federation 4 Future Internet Research and Experimentation Plus}
\newacronym{fir}{FIR}{Finite Impulse Response}
\newacronym{fit}{FIT}{Future \acrlong{iot}}
\newacronym{fpga}{FPGA}{Field Programmable Gate Array}
\newacronym{fr1}{FR1}{Frequency Range 1}
\newacronym{fr2}{FR2}{Frequency Range 2}
\newacronym{fs}{FS}{Fast Switching}
\newacronym{fscc}{FSCC}{Flow Sharing Congestion Control}
\newacronym{ftp}{FTP}{File Transfer Protocol}
\newacronym{fw}{FW}{Flow Window}
\newacronym{ge}{GE}{Gaussian Elimination}
\newacronym{gnb}{gNB}{Next Generation Node Base}
\newacronym{gop}{GOP}{Group of Pictures}
\newacronym{gpr}{GPR}{Gaussian Process Regressor}
\newacronym{gpu}{GPU}{Graphics Processing Unit}
\newacronym{gtp}{GTP}{GPRS Tunneling Protocol}
\newacronym{gtpc}{GTP-C}{GPRS Tunnelling Protocol Control Plane}
\newacronym{gtpu}{GTP-U}{GPRS Tunnelling Protocol User Plane}
\newacronym{gtpv2c}{GTPv2-C}{\gls{gtp} v2 - Control}
\newacronym{gw}{GW}{Gateway}
\newacronym{harq}{HARQ}{Hybrid Automatic Repeat reQuest}
\newacronym{hetnet}{HetNet}{Heterogeneous Network}
\newacronym{hh}{HH}{Hard Handover}
\newacronym{hol}{HOL}{Head-of-Line}
\newacronym{hqf}{HQF}{Highest-quality-first}
\newacronym{hss}{HSS}{Home Subscription Server}
\newacronym{http}{HTTP}{HyperText Transfer Protocol}
\newacronym{ia}{IA}{Initial Access}
\newacronym{iab}{IAB}{Integrated Access and Backhaul}
\newacronym{ic}{IC}{Incident Command}
\newacronym{ietf}{IETF}{Internet Engineering Task Force}
\newacronym{imsi}{IMSI}{International Mobile Subscriber Identity}
\newacronym{imt}{IMT}{International Mobile Telecommunication}
\newacronym{iot}{IoT}{Internet of Things}
\newacronym{ip}{IP}{Internet Protocol}
\newacronym{itu}{ITU}{International Telecommunication Union}
\newacronym{kpi}{KPI}{Key Performance Indicator}
\newacronym{kpm}{KPM}{Key Performance Measurement}
\newacronym{kvm}{KVM}{Kernel-based Virtual Machine}
\newacronym{los}{LOS}{Line-of-Sight}
\newacronym{lsm}{LSM}{Link-to-System Mapping}
\newacronym{lstm}{LSTM}{Long Short Term Memory}
\newacronym{lte}{LTE}{Long Term Evolution}
\newacronym{lxc}{LXC}{Linux Container}
\newacronym{m2m}{M2M}{Machine to Machine}
\newacronym{mac}{MAC}{Medium Access Control}
\newacronym{manet}{MANET}{Mobile Ad Hoc Network}
\newacronym{mano}{MANO}{Management and Orchestration}
\newacronym{mc}{MC}{Multi-Connectivity}
\newacronym{mcc}{MCC}{Mobile Cloud Computing}
\newacronym{mchem}{MCHEM}{Massive Channel Emulator}
\newacronym{mcs}{MCS}{Modulation and Coding Scheme}
\newacronym{mec}{MEC}{Multi-access Edge Computing}
\newacronym{mec2}{MEC}{Mobile Edge Cloud}
\newacronym{mfc}{MFC}{Mobile Fog Computing}
\newacronym{mgen}{MGEN}{Multi-Generator}
\newacronym{mi}{MI}{Mutual Information}
\newacronym{mib}{MIB}{Master Information Block}
\newacronym{miesm}{MIESM}{Mutual Information Based Effective SINR}
\newacronym{mimo}{MIMO}{Multiple Input, Multiple Output}
\newacronym{ml}{ML}{Machine Learning}
\newacronym{mlr}{MLR}{Maximum-local-rate}
\newacronym[plural=\gls{mme}s,firstplural=Mobility Management Entities (MMEs)]{mme}{MME}{Mobility Management Entity}
\newacronym{mmtc}{mMTC}{Massive Machine-Type Communications}
\newacronym{mmwave}{mmWave}{millimeter wave}
\newacronym{mpdccp}{MP-DCCP}{Multipath Datagram Congestion Control Protocol}
\newacronym{mptcp}{MPTCP}{Multipath TCP}
\newacronym{mr}{MR}{Maximum Rate}
\newacronym{mrdc}{MR-DC}{Multi \gls{rat} \gls{dc}}
\newacronym{mse}{MSE}{Mean Square Error}
\newacronym{mss}{MSS}{Maximum Segment Size}
\newacronym{mt}{MT}{Mobile Termination}
\newacronym{mtd}{MTD}{Machine-Type Device}
\newacronym{mtu}{MTU}{Maximum Transmission Unit}
\newacronym{mumimo}{MU-MIMO}{Multi-user \gls{mimo}}
\newacronym{mvno}{MVNO}{Mobile Virtual Network Operator}
\newacronym{nalu}{NALU}{Network Abstraction Layer Unit}
\newacronym{nas}{NAS}{Network Attached Storage}
\newacronym{nat}{NAT}{Network Address Translation}
\newacronym{nbiot}{NB-IoT}{Narrow Band IoT}
\newacronym{nfv}{NFV}{Network Function Virtualization}
\newacronym{nfvi}{NFVI}{Network Function Virtualization Infrastructure}
\newacronym{ni}{NI}{Network Interfaces}
\newacronym{nic}{NIC}{Network Interface Card}
\newacronym{nlos}{NLOS}{Non-Line-of-Sight}
\newacronym{now}{NOW}{Non Overlapping Window}
\newacronym{nsm}{NSM}{Network Service Mesh}
\newacronym[type=hidden]{nr}{NR}{New Radio}
\newacronym{nextg}{NextG}{Next Generation}
\newacronym{nrf}{NRF}{Network Repository Function}
\newacronym{nsa}{NSA}{Non Stand Alone}
\newacronym{nse}{NSE}{Network Slicing Engine}
\newacronym{nssf}{NSSF}{Network Slice Selection Function}
\newacronym{o2i}{O2I}{Outdoor to Indoor}
\newacronym{oai}{OAI}{OpenAirInterface}
\newacronym{oaicn}{OAI-CN}{\gls{oai} \acrlong{cn}}
\newacronym{oairan}{OAI-RAN}{\acrlong{oai} \acrlong{ran}}
\newacronym{oam}{OAM}{Operations, Administration and Maintenance}
\newacronym{ofdm}{OFDM}{Orthogonal Frequency Division Multiplexing}
\newacronym{olia}{OLIA}{Opportunistic Linked Increase Algorithm}
\newacronym{omec}{OMEC}{Open Mobile Evolved Core}
\newacronym{onap}{ONAP}{Open Network Automation Platform}
\newacronym{onf}{ONF}{Open Networking Foundation}
\newacronym{onos}{ONOS}{Open Networking Operating System}
\newacronym{oom}{OOM}{\gls{onap} Operations Manager}
\newacronym{opnfv}{OPNFV}{Open Platform for \gls{nfv}}
\newacronym[type=hidden]{oran}{O-RAN}{Open \gls{ran}}
\newacronym{orbit}{ORBIT}{Open-Access Research Testbed for Next-Generation Wireless Networks}
\newacronym{os}{OS}{Operating System}
\newacronym{osm2}{OSM}{Open Street Map}
\newacronym{oss}{OSS}{Operations Support System}
\newacronym{pa}{PA}{Position-aware}
\newacronym{pase}{PASE}{Prioritization, Arbitration, and Self-adjusting Endpoints}
\newacronym{pawr}{PAWR}{Platforms for Advanced Wireless Research}
\newacronym{pbch}{PBCH}{Physical Broadcast Channel}
\newacronym{pcef}{PCEF}{Policy and Charging Enforcement Function}
\newacronym{pcfich}{PCFICH}{Physical Control Format Indicator Channel}
\newacronym{pcrf}{PCRF}{Policy and Charging Rules Function}
\newacronym{pdcch}{PDCCH}{Physical Downlink Control Channel}
\newacronym{pdcp}{PDCP}{Packet Data Convergence Protocol}
\newacronym{pdsch}{PDSCH}{Physical Downlink Shared Channel}
\newacronym{pdu}{PDU}{Packet Data Unit}
\newacronym{pf}{PF}{Proportional Fair}
\newacronym{pgw}{PGW}{Packet Gateway}
\newacronym{phich}{PHICH}{Physical Hybrid ARQ Indicator Channel}
\newacronym{phy}{PHY}{Physical}
\newacronym{pl}{PL}{Path Loss}
\newacronym{pmch}{PMCH}{Physical Multicast Channel}
\newacronym{pmi}{PMI}{Precoding Matrix Indicators}
\newacronym{powder}{POWDER}{Platform for Open Wireless Data-driven Experimental Research}
\newacronym{ppo}{PPO}{Proximal Policy Optimization}
\newacronym{ppp}{PPP}{Poisson Point Process}
\newacronym{prach}{PRACH}{Physical Random Access Channel}
\newacronym{prb}{PRB}{Physical Resource Block}
\newacronym{psnr}{PSNR}{Peak Signal to Noise Ratio}
\newacronym{pss}{PSS}{Primary Synchronization Signal}
\newacronym{pucch}{PUCCH}{Physical Uplink Control Channel}
\newacronym{pusch}{PUSCH}{Physical Uplink Shared Channel}
\newacronym{qam}{QAM}{Quadrature Amplitude Modulation}
\newacronym{qci}{QCI}{\gls{qos} Class Identifier}
\newacronym{qoe}{QoE}{Quality of Experience}
\newacronym{qos}{QoS}{Quality of Service}
\newacronym{quic}{QUIC}{Quick UDP Internet Connections}
\newacronym{rach}{RACH}{Random Access Channel}
\newacronym{ran}{RAN}{Radio Access Network}
\newacronym[firstplural=Radio Access Technologies (RATs)]{rat}{RAT}{Radio Access Technology}
\newacronym{rbg}{RBG}{Resource Block Group}
\newacronym{rcn}{RCN}{Research Coordination Network}
\newacronym{rc}{RC}{RAN Control}
\newacronym{rec}{REC}{Radio Edge Cloud}
\newacronym{red}{RED}{Random Early Detection}
\newacronym{renew}{RENEW}{Reconfigurable Eco-system for Next-generation End-to-end Wireless}
\newacronym{rf}{RF}{Radio Frequency}
\newacronym{rfc}{RFC}{Request for Comments}
\newacronym{rfr}{RFR}{Random Forest Regressor}
\newacronym{ric}{RIC}{RAN Intelligent Controller}
\newacronym{rlc}{RLC}{Radio Link Control}
\newacronym{rlf}{RLF}{Radio Link Failure}
\newacronym{rlnc}{RLNC}{Random Linear Network Coding}
\newacronym{rmr}{RMR}{RIC Message Router}
\newacronym{rmse}{RMSE}{Root Mean Squared Error}
\newacronym{rnis}{RNIS}{Radio Network Information Service}
\newacronym{rr}{RR}{Round Robin}
\newacronym{rrc}{RRC}{Radio Resource Control}
\newacronym{rrm}{RRM}{Radio Resource Management}
\newacronym{rru}{RRU}{Remote Radio Unit}
\newacronym{rs}{RS}{Remote Server}
\newacronym{rsrp}{RSRP}{Reference Signal Received Power}
\newacronym{rsrq}{RSRQ}{Reference Signal Received Quality}
\newacronym{rss}{RSS}{Received Signal Strength}
\newacronym{rssi}{RSSI}{Received Signal Strength Indicator}
\newacronym{rt}{RT}{Real-time}
\newacronym{rtt}{RTT}{Round Trip Time}
\newacronym{ru}{RU}{Radio Unit}
\newacronym{rw}{RW}{Receive Window}
\newacronym{rx}{RX}{Receiver}
\newacronym{s1ap}{S1AP}{S1 Application Protocol}
\newacronym{sa}{SA}{standalone}
\newacronym{sack}{SACK}{Selective Acknowledgment}
\newacronym{sap}{SAP}{Service Access Point}
\newacronym{sc2}{SC2}{Spectrum Collaboration Challenge}
\newacronym{scef}{SCEF}{Service Capability Exposure Function}
\newacronym{sch}{SCH}{Secondary Cell Handover}
\newacronym{scoot}{SCOOT}{Split Cycle Offset Optimization Technique}
\newacronym{sctp}{SCTP}{Stream Control Transmission Protocol}
\newacronym{sdap}{SDAP}{Service Data Adaptation Protocol}
\newacronym{sdk}{SDK}{Software Development Kit}
\newacronym{sdm}{SDM}{Space Division Multiplexing}
\newacronym{sdma}{SDMA}{Spatial Division Multiple Access}
\newacronym{sdn}{SDN}{Software-defined Networking}
\newacronym{sdr}{SDR}{Software-defined Radio}
\newacronym{seba}{SEBA}{SDN-Enabled Broadband Access}
\newacronym{sgsn}{SGSN}{Serving GPRS Support Node}
\newacronym{sgw}{SGW}{Service Gateway}
\newacronym{si}{SI}{Study Item}
\newacronym{sib}{SIB}{Secondary Information Block}
\newacronym{sinr}{SINR}{Signal to Interference plus Noise Ratio}
\newacronym{sip}{SIP}{Session Initiation Protocol}
\newacronym{siso}{SISO}{Single Input, Single Output}
\newacronym{sla}{SLA}{Service Level Agreement}
\newacronym{sm}{SM}{Service Model}
\newacronym{smf}{SMF}{Session Management Function}
\newacronym{smo}{SMO}{Service Management and Orchestration}
\newacronym{sms}{SMS}{Short Message Service}
\newacronym{smsgmsc}{SMS-GMSC}{\gls{sms}-Gateway}
\newacronym{snr}{SNR}{Signal-to-Noise-Ratio}
\newacronym{son}{SON}{Self-Organizing Network}
\newacronym{sptcp}{SPTCP}{Single Path TCP}
\newacronym{srb}{SRB}{Service Radio Bearer}
\newacronym{srn}{SRN}{Standard Radio Node}
\newacronym{srs}{SRS}{Sounding Reference Signal}
\newacronym{ss}{SS}{Synchronization Signal}
\newacronym{sss}{SSS}{Secondary Synchronization Signal}
\newacronym{st}{ST}{Spanning Tree}
\newacronym{svc}{SVC}{Scalable Video Coding}
\newacronym{tb}{TB}{Transport Block}
\newacronym{tcp}{TCP}{Transmission Control Protocol}
\newacronym{tdd}{TDD}{Time Division Duplexing}
\newacronym{tdl}{TDL}{Tapped Delay Line}
\newacronym{tdm}{TDM}{Time Division Multiplexing}
\newacronym{tdma}{TDMA}{Time Division Multiple Access}
\newacronym{tfl}{TfL}{Transport for London}
\newacronym{tfrc}{TFRC}{TCP-Friendly Rate Control}
\newacronym{tft}{TFT}{Traffic Flow Template}
\newacronym{tgen}{TGEN}{Traffic Generator}
\newacronym{tip}{TIP}{Telecom Infra Project}
\newacronym{tm}{TM}{Transparent Mode}
\newacronym{to}{TO}{Telco Operator}
\newacronym{tr}{TR}{Technical Report}
\newacronym{trp}{TRP}{Transmitter Receiver Pair}
\newacronym{ts}{TS}{Technical Specification}
\newacronym{tti}{TTI}{Transmission Time Interval}
\newacronym{ttt}{TTT}{Time-to-Trigger}
\newacronym{tx}{TX}{Transmitter}
\newacronym{uas}{UAS}{Unmanned Aerial System}
\newacronym{uav}{UAV}{Unmanned Aerial Vehicle}
\newacronym{udm}{UDM}{Unified Data Management}
\newacronym{udp}{UDP}{User Datagram Protocol}
\newacronym{udr}{UDR}{Unified Data Repository}
\newacronym{ue}{UE}{User Equipment}
\newacronym{uhd}{UHD}{\gls{usrp} Hardware Driver}
\newacronym{ul}{UL}{Uplink}
\newacronym{um}{UM}{Unacknowledged Mode}
\newacronym{umi}{UMi}{Urban Micro}
\newacronym{uml}{UML}{Unified Modeling Language}
\newacronym{upa}{UPA}{Uniform Planar Array}
\newacronym{upf}{UPF}{User Plane Function}
\newacronym{urllc}{URLLC}{Ultra Reliable and Low Latency Communications}
\newacronym{usa}{U.S.}{United States}
\newacronym{usim}{USIM}{Universal Subscriber Identity Module}
\newacronym{usrp}{USRP}{Universal Software Radio Peripheral}
\newacronym{utc}{UTC}{Urban Traffic Control}
\newacronym{vim}{VIM}{Virtualization Infrastructure Manager}
\newacronym{vm}{VM}{Virtual Machine}
\newacronym{vnf}{VNF}{Virtual Network Function}
\newacronym{volte}{VoLTE}{Voice over \gls{lte}}
\newacronym{voltha}{VOLTHA}{Virtual OLT HArdware Abstraction}
\newacronym{vr}{VR}{Virtual Reality}
\newacronym{vran}{vRAN}{Virtualized \gls{ran}}
\newacronym{vss}{VSS}{Video Streaming Server}
\newacronym{wbf}{WBF}{Wired Bias Function}
\newacronym{wf}{WF}{Waterfilling}
\newacronym{wg}{WG}{Working Group}
\newacronym{wi}{WI}{Wireless InSite}
\newacronym{wlan}{WLAN}{Wireless Local Area Network}
\newacronym{osm}{OSM}{Open Source \gls{nfv} Management and Orchestration}
\newacronym{pnf}{PNF}{Physical Network Function}
\newacronym{mtc}{MTC}{Machine-type Communications}
\newacronym{mns}{MnS}{Management Services}
\newacronym{ves}{VES}{\gls{vnf} Event Stream}
\newacronym{ei}{EI}{Enrichment Information}
\newacronym{fh}{FH}{Fronthaul}
\newacronym{fft}{FFT}{Fast Fourier Transform}
\newacronym{laa}{LAA}{Licensed-Assisted Access}
\newacronym{plfs}{PLFS}{Physical Layer Frequency Signals}
\newacronym{ptp}{PTP}{Precision Time Protocol}
\newacronym{cbrs}{CBRS}{Citizen Broadband Radio Service}
\newacronym{otic}{OTIC}{Open Testing and Integration Center}
\newacronym{sba}{SBA}{Service-Based Architecture}
\newacronym{cif}{CI}{cyberinfrastructure}
\newacronym{sonic}{SONiC}{Software for Open Networking in the Cloud}
\newacronym{ocp}{OCP}{Open Compute Project}
\newacronym{snmp}{SNMP}{Simple Network Management Protocol}
\newacronym{raid}{RAID}{redundant array of independent disks}
\newacronym{nfs}{NFS}{Network File Storage}
\newacronym{ci}{CI}{Continuous Integration}
\newacronym{cd}{CD}{Continuous Deployment}
\newacronym{dtn}{DTN}{Data Transfer Node}

\newacronym{dns}{DNS}{Domain Name Service}
\newacronym{nrpe}{NRPE}{Nagios Remote Plugin Executor}
\newacronym{ldap}{LDAP}{Lightweight Directory Access Protocol}
\newacronym{lan}{LAN}{Local Area Network}
\newacronym{vlan}{VLAN}{Virtual LAN}

\newacronym{ipmi}{IPMI}{Intelligent Platform Management Interface}
\newacronym{tor}{ToR}{Top-of-the-Rack}
\newacronym{lmn}{LMN}{Large Memory Node}
\newacronym{bgp}{BGP}{Border Gateway Protocol}
\newacronym{dhcp}{DHCP}{Dynamic Host Configuration Protocol}
\newacronym{vrf}{VRF}{Virtual Routing and Forwarding}
\newacronym{vpn}{VPN}{Virtual Private Network}
\newacronym{rma}{RMA}{Return Merchandise Authorization}
\newacronym{hpc}{HPC}{High Performance Compute}

\newacronym{nu}{NU}{Northeastern University}
\newacronym{asic}{ASIC}{Application-specific Integrated Circuit}
\newacronym{rdma}{RDMA}{Remote Direct Memory Access}
\newacronym{roce}{RoCE}{RDMA over Converged Ethernet}
\newacronym{ovs}{OVS}{Open vSwitch}
\newacronym{frr}{FRR}{Free Range Routing}
\newacronym{ups}{UPS}{Uninterruptible Power Supply}

\newacronym{ntia}{NTIA}{National Telecommunications and Information Administration}
\newacronym{pii}{PII}{Personal and Identifiable Information}
\newacronym{irb}{IRB}{Institutional Review Board}
\newacronym{doi}{DOI}{Digital Object Identifier}

\newacronym{sdo}{SDO}{Standard-Development Organization}
\newacronym{ose}{OSE}{Open Source Ecosystem}
\newacronym{osc}{OSC}{O-RAN Software Community}
\newacronym{dop}{DOP}{Director of Operations}
\newacronym{pm}{PM}{Program Manager}
\newacronym{excom}{EXCOM}{Executive Committee}
\newacronym{iiot}{IIoT}{Industrial \gls{iot}}
\newacronym{lf}{LF}{Linux Foundation}

\newacronym{wiot}{WIoT}{Institute for the Wireless Internet of Things}
\newacronym{rl}{RL}{Reinforcement Learning}
\newacronym{drl}{DRL}{Deep Reinforcement Learning}

\newacronym{nofo}{NOFO}{Notice of Funding Opportunity}

\newacronym{onr}{ONR}{Office of Naval Research}
\newacronym{afosr}{AFOSR}{Air Force Office of Scientific Research}
\newacronym{afrl}{AFRL}{Air Force Research Laboratory}
\newacronym{arl}{ARL}{Army Research Laboratory}

\newacronym{arc}{ARC}{Aerial Research Cloud}
\newacronym{cast}{CaST}{Channel emulation scenario generator and Sounder Toolchain}

\newacronym{mno}{MNO}{Mobile Network Operator}
\newacronym{ct}{CT}{Continuous Testing}
\newacronym{oci}{OCI}{Open Container Initiative}

\newacronym{xai}{XAI}{Explainable AI}
\newacronym{esc}{ESC}{Environmental Sensing Capability}
\newacronym{sas}{SAS}{Spectrum Access System}

\newacronym{rem}{REM}{Random Ensemble Mixture}
\newacronym{ns3}{ns-3}{Network Simulator 3}
\tikzstyle{startstop} = [rectangle, rounded corners, minimum width=2cm, minimum height=0.5cm,text centered, draw=black]
\tikzstyle{io} = [trapezium, trapezium left angle=70, trapezium right angle=110, minimum width=3cm, minimum height=1cm, text centered, draw=black]
\tikzstyle{process} = [rectangle, minimum width=2cm, minimum height=0.5cm, text centered, draw=black, alignb=center]
\tikzstyle{decision} = [ellipse, minimum width=2cm, minimum height=1cm, text centered, draw=black]
\tikzstyle{arrow} = [thick,<->,>=stealth]
\tikzstyle{line} = [thick,>=stealth]
\tikzstyle{darrow} = [thick,<->,>=stealth,dashed]
\tikzstyle{sarrow} = [thick,->,>=stealth]
\tikzstyle{larrow} = [line width=0.3mm,dashdotted,->,>=stealth]
\tikzstyle{llarrow} = [line width=0.1mm,->,>=stealth]

\makeatletter
\def\grd@save@target#1{%
  \def\grd@target{#1}}
\def\grd@save@start#1{%
  \def\grd@start{#1}}
\tikzset{
  grid with coordinates/.style={
    to path={%
      \pgfextra{%
        \edef\grd@@target{(\tikztotarget)}%
        \tikz@scan@one@point\grd@save@target\grd@@target\relax
        \edef\grd@@start{(\tikztostart)}%
        \tikz@scan@one@point\grd@save@start\grd@@start\relax
        \draw[minor help lines] (\tikztostart) grid (\tikztotarget);
        \draw[major help lines] (\tikztostart) grid (\tikztotarget);
        \grd@start
        \pgfmathsetmacro{\grd@xa}{\the\pgf@x/1cm}
        \pgfmathsetmacro{\grd@ya}{\the\pgf@y/1cm}
        \grd@target
        \pgfmathsetmacro{\grd@xb}{\the\pgf@x/1cm}
        \pgfmathsetmacro{\grd@yb}{\the\pgf@y/1cm}
        \pgfmathsetmacro{\grd@xc}{\grd@xa + \pgfkeysvalueof{/tikz/grid with coordinates/major step x}}
        \pgfmathsetmacro{\grd@yc}{\grd@ya + \pgfkeysvalueof{/tikz/grid with coordinates/major step y}}
        \foreach \x in {\grd@xa,\grd@xc,...,\grd@xb}
        \node[anchor=north] at (\x,\grd@ya) {\pgfmathprintnumber{\x}};
        \foreach \y in {\grd@ya,\grd@yc,...,\grd@yb}
        \node[anchor=east] at (\grd@xa,\y) {\pgfmathprintnumber{\y}};
      }
    }
  },
  minor help lines/.style={
    help lines,
    gray,
    line cap =round,
    xstep=\pgfkeysvalueof{/tikz/grid with coordinates/minor step x},
    ystep=\pgfkeysvalueof{/tikz/grid with coordinates/minor step y}
  },
  major help lines/.style={
    help lines,
    line cap =round,
    line width=\pgfkeysvalueof{/tikz/grid with coordinates/major line width},
    xstep=\pgfkeysvalueof{/tikz/grid with coordinates/major step x},
    ystep=\pgfkeysvalueof{/tikz/grid with coordinates/major step y}
  },
  grid with coordinates/.cd,
  minor step x/.initial=.5,
  minor step y/.initial=.2,
  major step x/.initial=1,
  major step y/.initial=1,
  major line width/.initial=1pt,
}
\makeatother

\usepackage{dblfloatfix}

\begin{document}
\title{Design and Evaluation of Deep Reinforcement Learning for Energy Saving in Open RAN}


\author{\IEEEauthorblockN{Matteo Bordin\IEEEauthorrefmark{1}, Andrea Lacava\IEEEauthorrefmark{1}\IEEEauthorrefmark{2}, Michele Polese\IEEEauthorrefmark{1}, Sai Satish\IEEEauthorrefmark{3}, Manoj AnanthaSwamy Nittoor\IEEEauthorrefmark{3},\\ 
Rajarajan Sivaraj\IEEEauthorrefmark{3}, Francesca Cuomo\IEEEauthorrefmark{2}, Tommaso Melodia\IEEEauthorrefmark{1}}
\IEEEauthorblockN{\IEEEauthorrefmark{1}Institute for the Wireless Internet of Things, Northeastern University, Boston, MA, USA\\
\IEEEauthorrefmark{2}Sapienza University of Rome, Rome, Italy, \IEEEauthorrefmark{3}Mavenir, Richardson, TX, USA\\
Email: \{{bordin.m, lacava.a, m.polese, t.melodia\}@northeastern.edu}\\
Email: \{{sai.satish, manoj.nittoor, rajarajan.sivaraj\}@mavenir.com}, \{{francesca.cuomo\}@uniroma1.it}\\
}}

\maketitle

\begin{abstract}
Next-generation wireless systems, already widely deployed, are expected to become even more prevalent in the future, representing challenges in both environmental and economic terms.
This paper focuses on improving the energy efficiency of intelligent and programmable Open \gls{ran} systems through the near-real-time dynamic activation and deactivation of \gls{bs} \gls{rf} frontends using \gls{drl} algorithms, i.e., \gls{ppo} and \gls{dqn}. 
These algorithms run on the \glspl{ric}, part of the Open \gls{ran} architecture, and are designed to make optimal network-level decisions based on historical data without compromising stability and performance. 
We leverage a rich set of \glspl{kpm}, serving as state for the \gls{drl}, to create a comprehensive representation of the \gls{ran}, alongside a set of actions that correspond to some control exercised on the RF frontend.
We extend ns-O-RAN, an open-source, realistic simulator for 5G and Open RAN built on ns-3, to conduct an extensive data collection campaign. This enables us to train the agents offline with over 300,000 data points and subsequently evaluate the performance of the trained models. Results show that \gls{drl} agents improve energy efficiency by adapting to network conditions while minimally impacting the user experience.
Additionally, we explore the trade-off between throughput and energy consumption offered by different \gls{drl} agent designs.
\end{abstract}
\begin{picture}(0,0)(0,-450)
\put(0,0){
    \put(0,0){\small \shortstack[l]{\qquad \quad This paper has been accepted to IEEE CCNC 2025. If you wish to cite this work, please use the following reference:\\ \qquad \quad \quad \quad
    M. Bordin, A. Lacava, M. Polese, S. Satish, M. AnanthaSwamy Nittoor, R. Sivaraj, F. Cuomo, and T. Melodia, \\ \qquad \quad“Design and Evaluation of Deep Reinforcement Learning for Energy Saving in Open RAN,” in Proc. IEEE CCNC 2025.}}
}
\end{picture}

\glsresetall
\glsunset{ns3}
\glsunset{nr}
\glsunset{lte}


\section{Introduction}
\label{sec:intro}
The \gls{ran} in \gls{5g} systems is becoming increasingly complex due to the diverse range of use cases and the rising demand for data traffic. Next-generation wireless networks face a dual challenge: they must address growing data demands while also managing the significant energy consumption associated with their operation~\cite{9678321}.
As \gls{5g} \glspl{bs} are deployed at unprecedented densities to ensure widespread coverage and high-speed connectivity, the energy required to power this infrastructure becomes increasingly onerous. 
As a result, there is a critical need for innovative solutions that can intelligently manage energy consumption in \gls{5g} networks while maintaining optimal performance.
Today, traditional \gls{ran} \glspl{bs} are based on monolithic network systems and proprietary hardware~\cite{9678321,salem:tel-02500618,7060678}. Therefore manual intervention is needed for reconfiguration, rgy-saving plans. Open \gls{ran}, defined by the O-RAN ALLIANCE, breaks away from this traditional approach by applying the principles of disaggregation, openness, virtualization and programmability. 
In Open \gls{ran}, the classical \gls{bs} is disaggregated: interfaces between different logical nodes are open and standardized to achieve multi-vendor interoperability. Moreover, standard \gls{ran} operations are virtualized and deployed on white box hardware~\cite{polese2023understanding}, enabling dynamic infrastructure reconfigurations. This software-based approach allows the integration of xApps, which are algorithms designed to manage the \gls{ran} in real-time. xApps interact with the \gls{ran} through the E2 interface, enabling  efficient \gls{rrm} decisions~\cite{9839628}.


The main contribution of this paper is the development of a xApp focused on reducing the energy consumption of the \gls{ran} through dynamic activation and deactivation of the \gls{rf} frontends associated with specific cells which are responsible for 65\% of the \gls{ran} power consumption~\cite{9678321}.   
By leveraging real-time data analytics, \gls{drl} and a global point of view on the network, the energy-saving xApp can intelligently control the activation and deactivation of \gls{5g} cells to optimize energy consumption while maintaining \gls{qos} requirements. Specifically, we design and profile multiple energy-saving \gls{drl} models to explore the trade-off between energy-saving and throughput.
Another contribution of this paper is the introduction of energy-saving functionalities in ns-O-RAN, a ns-3 open-source module that models a 5G RAN with O-RAN components, providing a platform to develop and test xApps, allowing a realistic performance evaluation with 3GPP-based channel models and protocols~\cite{lacava2023ns}.
For the offline training of the \gls{drl} agents, a dataset is generated through a large-scale campaign on ns-3 with thousands of simulations and four different energy-saving heuristics from the literature~\cite{7060678}. 
We compare and test different agents based on \gls{ppo} and \gls{dqn}.
The agents maximize a reward function that combines throughput, energy consumption, coverage, and a cost factor associated with activating and deactivating cells too frequently, to avoid ping-pong effects on the users.
Our results show the existence of a trade-off between energy consumption and throughput, where the reduction of the first may necessitate a sacrifice of the latter. However, compared to heuristic methods, \gls{drl} agents offer greater flexibility. They can adapt to dynamic conditions, improving throughput while also reducing energy consumption.

The paper proceeds as follows: Section~\ref{sec:soa} reviews the state of the art, Section~\ref{sec:system} introduces the system architecture, and Section~\ref{sec:drl-agent-design} the optimization problem along with the \gls{drl} agent designs. Section~\ref{sec:experiments} details the experimental setup, including the \gls{ns3} scenario and baseline descriptions. Finally, Section~\ref{sec:results} and Section~\ref{sec:conclusions} present the final \gls{drl} model results and the overall conclusion.

\section{State of the Art}
\label{sec:soa}
The increasing data demand and capabilities of \glspl{ue} in \gls{5g} networks and beyond have raised significant concerns about power consumption, impacting the environment and operational costs. This has led to a continued focus on enhancing networks \gls{ee}, a topic of ongoing economic and social interest~\cite{7446253}. 
In~\cite{10200477}, a comprehensive survey is presented on the topic, highlighting different power consumption models and \gls{ee} techniques. Many works focus on developing efficient hardware to mitigate the impact of the \gls{mimo} arrays and of the \gls{ru} in general like~\cite{10008509} and~\cite{10380322}. 
Authors in ~\cite{7811130} review techniques to enhance \gls{ee} gains in massive \gls{mimo} systems, including hybrid configurations with millimeter wave and heterogeneous networks, highlighting both opportunities and challenges for maximizing \gls{ee} in future \gls{5g} deployments. 
Authors of ~\cite{SHUVO2021102986} and \cite{9772655}, pose energy consumption models based on scheduling algorithm inspired by biological genetics to minimize energy consumption. 
Another particular technique is cell zooming presented in \cite{DAHAL2022100040}: it optimizes energy consumption by dynamically adjusting the number of active base stations and their transmit power according to traffic volume while ensuring quality of service.
Other approaches start from solutions developed for LTE networks: \cite{9112193}, \cite{9246536} and \cite{9931944} demonstrate how LTE energy-saving methods can be easily adapted to 5G networks e.g. time and frequency domain and DRX mechanism. 
In the energy saving context, the Open \gls{ran} architecture can help to improve network scalability, orchestration, and \gls{rrm} leading to improve the energy efficiency of the \gls{ran} deployments.
Despite these promises, the impact of disaggregated architecture on energy consumption is still an open question and subject of interest. 
Other energy-saving solutions focus more on software approaches, exploring the possibilities enabled by the Open \gls{ran} Intelligent Control loops deployed on the \gls{ric}. These solutions utilize xApps for near-real-time operations and rApps for non-real-time periodic tasks \cite{polese2023understanding}.
In ~\cite{10437790}, authors design an energy consumption model and a pipeline for the deployment of ls{ml} models, to minimize their energy usage during their life cycle in the Open \gls{ran} context.
In contrast to previous works, our efforts are concentrated on developing a \gls{drl} agent to save energy in the network, able to interact with the \gls{ran} in near-real-time, using a control strategy involving dynamic activation and deactivation of the \glspl{gnb} \gls{rf} frontend. Unlike existing methods, \gls{drl} optimizes decisions with a global network view, learning and adapting to new patterns without the need for explicit programming for every scenario.
Finally, to create meaningful data to feed our \gls{drl} agents, we have implemented in ns-O-RAN three different baselines based on~\cite{7060678}, where the authors define two strategic and one random sleeping policies, with constraints on both wake-up times and coverage probabilities. A description of the heuristics implemented in ns-O-RAN is in Sec.~\ref{sec:experiments}.


\section{System overview}
\label{sec:system} 



\begin{figure}[t]
\centering
\includegraphics[width=0.9\columnwidth]{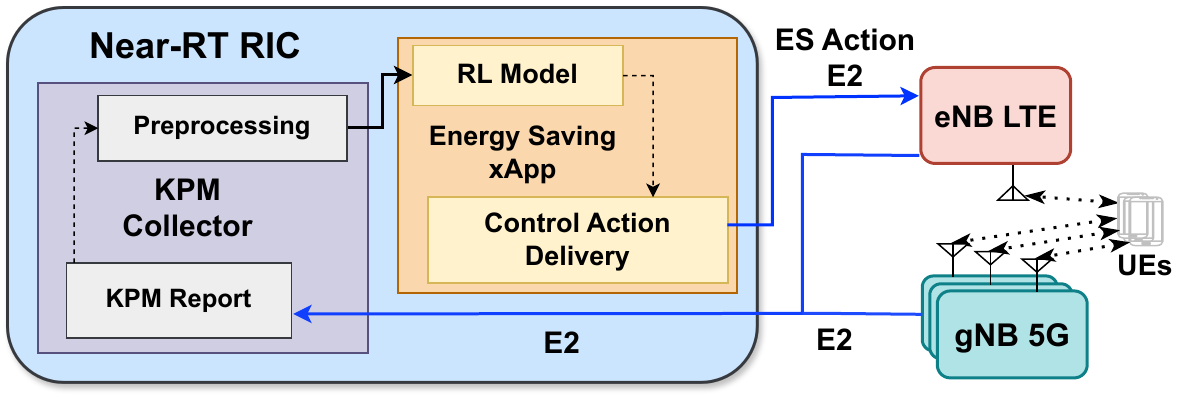}
\caption{System architecture}
\label{fig:SystArchit}
\end{figure}

The system architecture is shown in Fig.~\ref{fig:SystArchit}.
It consists of one \gls{lte} \gls{enb}, $N$ \gls{nr} \gls{5g} \gls{gnb} deployed as a \gls{nsa} network, and a set $U$ of \gls{5g} \glspl{ue}. 
\glspl{ue} have heterogeneous types of data traffic, as detailed in Sec.~\ref{sec:experiments}.
In addition, a near-real-time \gls{ric} is connected to each \gls{lte} and \gls{nr} cell through the E2 interface. The E2 interface manages the exchange of network data and cell state control actions at near-real-time periodicity.
The near-real-time \gls{ric} is deployed at the edge of the \gls{ran} and features an E2 termination, a \gls{kpm} collector that gathers all the performance metrics from both the \gls{enb} and \gls{gnb} cells and the ES xApp to apply the control actions to each \gls{5g} cell. 
The xApp, running in the near-real-time \gls{ric}, generates control actions from the \gls{ric} to the E2 node to determine the activated or deactivated state of the cells.
The choice of using an xApp instead of a rApp is due to the need for a control mechanism that can quickly respond to changes in network conditions, as discussed next.


\textbf{RAN Control for Energy Efficiency and Sleep Modes.} 
The \gls{bs} can have different sleep modes, which correspond to the deactivation of different parts of the node, from the power amplifier to the full RF front end and up to the baseband processing unit. Different sleep modes are also mapped to different delays in deactivating and reactivating the components of the \gls{bs}~\cite{9678321,salem:tel-02500618}.
Sleep mode 1 is characterized by a duration of deactivation plus a reactivation time of 71 microseconds, and the power amplifier, some components of the digital \gls{bbu}, and the analog frontend can be deactivated. In sleep mode 2, which has a duration of deactivation plus reactivation of 1\,ms, additional components of the analog frontend can be switched off. In Sleep mode 3, which has a minimum duration of deactivation and activation of 10\,ms, the \gls{bs} can additionally deactivate all the digital \gls{bbu} processing, and almost all the analog frontend (except the clock generator). Finally, in sleep mode 4, which has a minimum duration of deactivation plus activation of 1\,s, only the wake-up functionalities are maintained.
The energy saving performed in this paper is mode 1, where deactivation and reactivation are executed as quickly as possible. This is particularly notable because the RF equipment (such as the power amplifier, transceivers, and cables) has been identified as the largest energy consumer within a BS, typically using about 65\% of the total BS energy. ~\cite{9678321}.

\section{DRL Agent Design}
\label{sec:drl-agent-design}
In this section, we formulate the optimization problem for the energy-saving xApp and discuss the algorithm design. 

We chose an offline \gls{drl} technique over online methods for two main reasons: the Open \gls{ran} specifications suggest not training online network models to reduce the risk of outages for \glspl{ue}~\cite{oran-wg2-ml}, and online training requires significant time for real-time data collection and agent-system interaction. 
Although offline learning does not directly use real-time transitions, its methodology focuses on optimizing policies based on historical data.
Details on the training dataset are discussed in Sec.~\ref{sec:experiments}.

\textbf{State.} To represent the state of the \gls{ran} used by the \gls{drl} model, the xApp collects \glspl{kpm} that are either exposed over the E2 interface by the $N$ NR base stations or computed in the xApp itself. 
\glspl{kpm} are functions representing the state of the RAN at time $t$ and are selected from all available states in the system due to their high correlation with the reward function.
With a slight abuse of notation, we represent $a_i(t)$ as the action applied at time $t$ to \gls{bs} $i \in \{1, \dots, N\}$, assuming the rest of the \gls{ran} configuration will not change. $a_i(t)$ indicates whether a \gls{bs} is active or not at time $t$.
We define the throughput $\rho_i$ as the number of bytes transmitted at the PDCP layer by cell $i$, and the energy consumption $\gamma_i$ of cell $i$ as 
\begin{equation}
    \gamma_{i}(a_i(t)) = EC_{i}(a_i(t)) \cdot P_{tx,i}
\end{equation}
where $EC_{i}$ and $P_{tx,i}$ are the total number of \gls{pdu} transmitted by cell $i$ and its transmit power, respectively. 
We also monitor $BsON(a_i(t))$ as the number of active \glspl{bs} in the scenario at time $t$.
The following 8 \glspl{kpm} are also calculated for each cell: the ratio between the throughput $\rho_i(\cdot)$ and the energy consumption $\gamma_{i}(\cdot)$; the number of \glspl{ue} in \gls{rlf} $\varsigma_i(\cdot)$, i.e., with a \gls{sinr} below -5\,dB; the percentage of \glspl{ue} in \gls{rlf} with respect to the total number of \glspl{ue} in $i$; the number and percentage of scheduled \glspl{prb}; the number of \gls{mac} \glspl{pdu} with a \gls{mcs} that uses 64-QAM; the number of bytes transmitted at the physical layer; and the cost $\delta_i(\cdot)$ to activate a single cell calculated as $0.9^{0.01 TD_i}$ where $TD_i$ is the time interval in which the cell remains active, measured in milliseconds. This last \gls{kpm} models an exponential decay function ensuring the cost is higher when there is a small time gap between action changes and lower cost when there is a longer gap. A total of 85 \glspl{kpm} ($12\cdot N +1$, with $N=7$) are collected in the simulation scenario described in Sec.~\ref{sec:experiments}. These KPMs serve as state for our \gls{drl} model, capturing the essential features of the environment that the model uses to make informed decisions.

\textbf{Reward function.} The target of the agent is to maximize the throughput of \glspl{ue} by finding an optimal trade-off between throughput and energy consumption and, at the same time, to avoid \glspl{rlf} of the individual \glspl{ue} and the same cell changing frequently its state. 
This is done by defining the reward function in Eq.~\eqref{eqn:es_problem_formulation} as a weighted combination of these \glspl{kpm} with weights $w_j, \; j \in \{1,\dots,4\}$ as follows:
%
%
\begin{align}
\max_{\mathbf{a}} \sum_{t=t_0}^{\infty}\sum_{i=1}^{N} & \{ w_1 \rho_{i}(a_i(t)) - w_2 \gamma_{i}(a_i(t)) - w_3 \varsigma_{i}(a_i(t)) \nonumber \\
& \qquad \qquad - \delta_{i}(a_i(t)) w_4 \label{eqn:es_problem_formulation} \} - w_2 BsON(a_i(t))  \tag{2}\\
\text{subject to} & \sum_{j=1}^{k} W_j=1 \tag{3}\\
    & a_i(t) \in A, \forall t = 0, \ldots, +\infty
    \tag{4}
\end{align}
As shown in Table~\ref{tab:RFweights}, we use different sets of values for the reward weights to dynamically assign different priorities to each component.
The table also reports the specific quantile normalization applied to each reward component. 
A quantile normalization transforms the data into intervals where each interval contains an equal number of interval boundaries called quantiles: it is usually applied when available data do not conform to a Gaussian or power-law distribution~\cite{googleNorm}. 
Finally, normal normalization in the range $[0, 1]$ has been applied to each remaining \gls{kpm} in the state vector.

\begin{table}[t]
\centering
\vspace{0.1in}
\renewcommand{\arraystretch}{1}
\caption{Reward function weights and normalization strategy.}
\label{tab:RFweights}
\begin{tabular}{llllll}
\toprule
Name & $w_1$ & $w_2$ & $w_3$ & $w_4$ & Quantile transformer\\
\midrule
PPO-1 & 0.51 & 0.19 & 0.2 & 0.1 & Normal\\
DQN & 0.4 & 0.4 & 0.1 & 0.1 & Uniform\\
PPO-2 & 0.4 & 0.4 & 0.1 & 0.1 & Uniform\\
PPO-3 & 0.2 & 0.4 & 0.2 & 0.2 & Uniform\\
PPO-4 & 0.4 & 0.32 & 0.18 & 0.1 & Uniform\\
PPO-5 & 0.45 & 0.2 & 0.25 & 0.1 & Normal\\
\bottomrule
\end{tabular}
\end{table}

\textbf{Action space.} Each action corresponds to a decision to either activate or deactivate a specific cell based on network conditions and maximizing the reward function. Since the environment involves 7 cells, the action space comprises 128 possible actions, while the \gls{drl} model action is represented as a 7-bit binary vector.


\textbf{\gls{drl} Agents Implementation.} The \gls{drl} models have been trained with two different algorithms: \gls{ppo} and \gls{dqn}.
For the \gls{ppo}, we used the implementation of Stable Baseline 3~\cite{raffin2021stable} with default hyperparameters except for the batch size equal to 256 and the entropy coefficient set in a range between 0.001 and 0.003 depending on the weights assigned to the reward function. 
The maximum number of timesteps has been set to 1200000 with an episode of 100 steps with the condition of an early stop if there was no improvement in the last four evaluations after a minimum of 10000 timesteps.
The model has been trained in a custom environment designed using the Gymnasium library~\cite{towers_gymnasium_2023}.
For the \gls{dqn}, we leverage the hybrid neural network design from~\cite{lacava2023programmable} with one 1D \gls{cnn} followed by three dense layers with sizes $[512, 512, 256]$. These configuration parameters are chosen to balance exploration and exploitation during training, ensuring stable and efficient learning. 
The training process involves iteratively updating the neural network parameters to minimize the discrepancy between predicted and targeted Q-values. 
To enhance stability and robustness in learning, we leverage the \gls{rem}-\gls{dqn} algorithm. Additionally, we integrate \gls{cql} into our framework to encourage conservative Q-value estimation and mitigate overestimation bias~\cite{NEURIPS2020_0d2b2061}.

\section{Simulation Setup}
\label{sec:experiments}
This paper's simulation environment is based on \gls{ns3}, an open-source simulator for complex network scenarios, allowing performance evaluation of networking algorithms and protocols. We specifically leverage the ns-O-RAN module~\cite{lacava2023ns}, which models the \gls{ran} and its connectivity to the near-real-time \gls{ric}. Baselines have been implemented to conduct extensive data collection for creating the training dataset for the \gls{drl} algorithms, as discussed next.

\subsection{Simulation Scenario}
\label{sec:scenario}

\begin{figure}[h]
\hspace*{-0.4cm}
\includegraphics[scale=0.46]{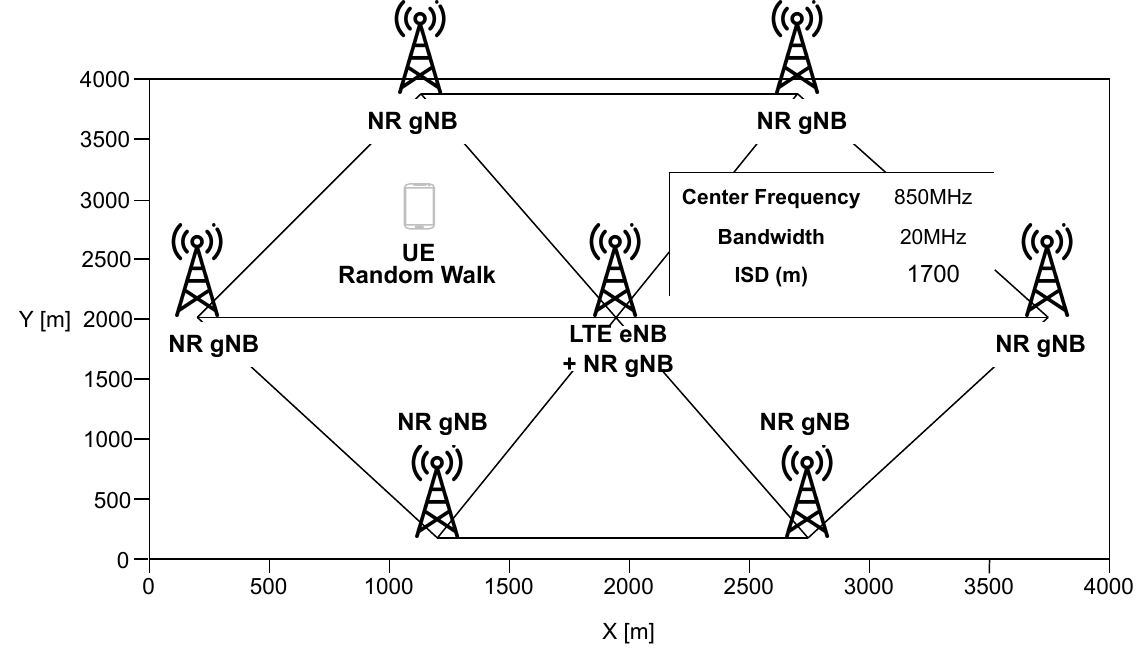}
\caption{Simulation scenario overview.}
\label{fig:ScenarioOne}
\end{figure}

We consider a Dense Urban Scenario described in~\cite{3gpp.38.913} with the \gls{3gpp} \gls{umi} Street Canyon channel model from~\cite{3gpp.38.901}. 
As shown in Fig.~\ref{fig:ScenarioOne}, this scenario features a \gls{nsa} \gls{5g} configuration, featuring a centrally located \gls{lte} \gls{enb} and one \gls{gnb} co-located with the \gls{lte} \gls{bs}, surrounded by seven additional \glspl{gnb}, each positioned 1700 meters from the center.
This scenario is deployed on the \gls{fr1}, using a center frequency of 850\,MHz with a bandwidth of 20\,MHz.
The scenario contains 9\,\glspl{ue} for each \gls{bs} for a total of 63\,\glspl{ue}. 
We implemented an \textit{uniform} positioning strategy for the \glspl{ue}, where, for each \gls{gnb}, 9 \glspl{ue} are randomly dropped uniformly in a disc centered in the \gls{enb} position and with a radius equal to the inter-site distance between \glspl{gnb}, and a \textit{non uniform} allocation strategy, where $\xi \in [1,2,3]$ \glspl{gnb} are excluded from the random positioning. 
The mobility model of the \glspl{ue} is a bidimensional random walk. Each \gls{ue} moves at a uniform speed between 2 m/s and 4 m/s. 
The user's downlink traffic is a mixture of four heterogeneous traffic models: 25\% of \glspl{ue} has a \gls{tcp} full-buffer traffic, with a data rate of 20\,Mbps; another 25\% has \gls{udp} bursty traffic with bit-rate averaging around 20\,Mbps, to simulate video streaming applications; the final 50\% has a \gls{tcp} bursty traffic where the first half has a bit-rate averaging 750\,kb/s, while the latter half 150\,kbps. The bursty traffic models feature on and off phases with a random exponential duration. 
The simulation time of this scenario is 10\,s and the periodicity of the control actions and messages to the \gls{ric} is 100\,ms.

\subsection{Energy-Saving Features and Baselines in ns-O-RAN} \label{sec:baselines}
To properly represent the optimization problem described in this paper, we have extended ns-O-RAN with the possibility of activating and deactivating a \gls{bs} frontend, thus modeling the control action for the energy-saving use case.  
In this new implementation, the primary \gls{enb} features a status map of the \glspl{gnb}, tracking whether a \gls{bs} is active or not.
Whenever the \gls{drl} agent in the xApp generates a control action, it is forwarded to the \gls{lte} \gls{enb}, which checks the status map, and if there is any update it performs the setup or the shutdown of the \gls{gnb} frontend. We consider a handover mechanism to automatically load balance the \glspl{ue} whenever a \gls{bs} is activated or deactivated, based on~\cite{polese2017improved}. 
The strategy for the decision of the target cells relies on a dynamic \gls{ttt}, which is decreased proportionally to the difference between the target and current cell \gls{sinr}.
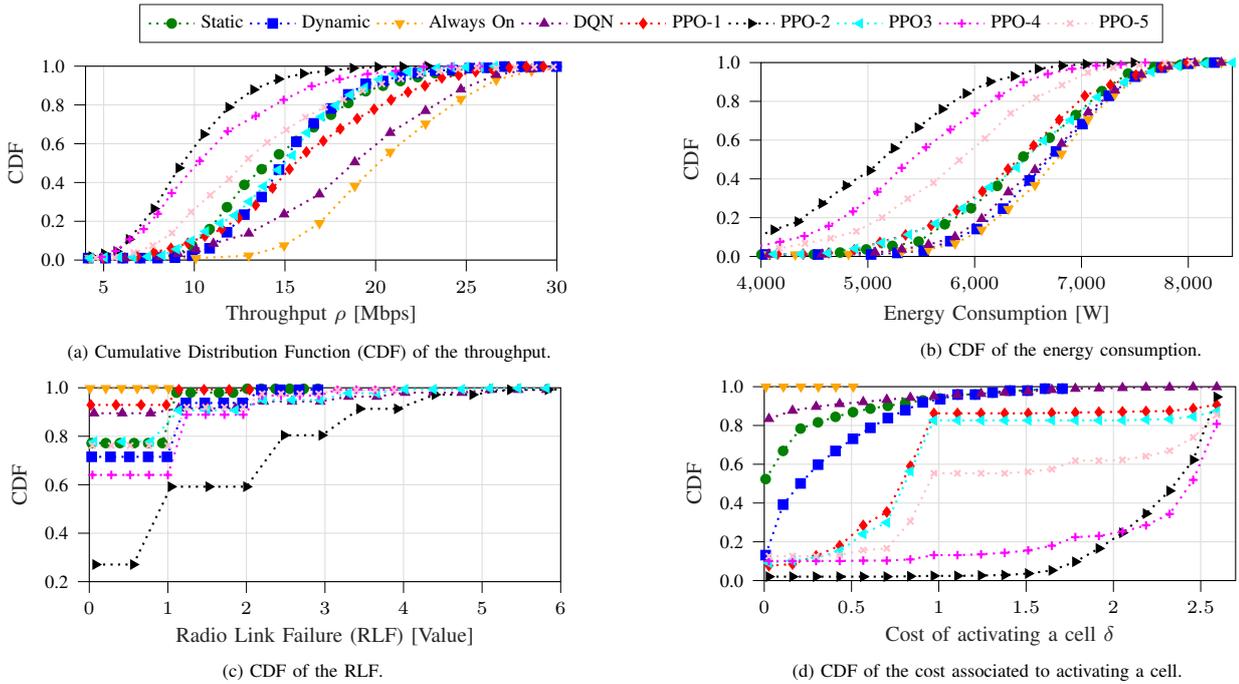
\begin{figure*}[t]
\vspace{0.1cm}
	\centering
	\hfill%
	\begin{subfigure}[t]{0.45\textwidth}
		\centering
		\setlength\fwidth{.8\textwidth}
		\setlength\fheight{.45\textwidth}
\begin{tikzpicture}
\pgfplotsset{every tick label/.append style={font=\scriptsize}}

\definecolor{color0}{rgb}{1,0.647058823529412,0} 
\definecolor{color1}{rgb}{0.501960784313725,0,0.501960784313725}
\definecolor{color2}{rgb}{0,1,1}
\definecolor{color3}{rgb}{1,0,1}
\definecolor{color4}{rgb}{1,0.752941176470588,0.796078431372549}

\begin{axis}[
width=0.95\fwidth,
height=0.7\fheight,
at={(0\fwidth,0\fheight)},
scale only axis,
tick align=outside,
tick pos=left,
x grid style={white!86.6666666666667!black},
xlabel={Throughput $\rho$ [Mbps]},
xlabel style={font=\footnotesize\color{white!15!black}},
xmajorgrids,
xmin=4, xmax=30,
xtick style={color=black},
y grid style={white!86.6666666666667!black},
ylabel={CDF},
ylabel style={font=\footnotesize\color{white!15!black}},
axis background/.style={fill=white},
ymajorgrids,
ymin=0, 
ymax=1,
ytick style={color=black},
ytick={-0.2,0,0.2,0.4,0.6,0.8,1,1.2},
yticklabels={−0.2,0.0,0.2,0.4,0.6,0.8,1.0,1.2},
legend style={font=\scriptsize,at={(1.2,1.12)},anchor=south,legend cell align=left,align=left,draw=white!15!black},
legend columns=9,
legend entries= {Static, Dynamic, Always On, DQN, PPO-1, PPO-2, PPO3, PPO-4, PPO-5},
xlabel shift={-2pt}
]
\addplot [thick, green!50.1960784313725!black, dotted, mark=*, mark size=1.6pt, mark options={solid}]
table {%
0.3189078 0.0101010101010101
1.2756312 0.0101010101010101
2.2323546 0.0101010101010101
3.189078 0.0101010101010101
4.1458014 0.0101010101010101
5.1025248 0.0101010101010101
6.0592482 0.0101010101010101
7.0159716 0.0101010101010101
7.972695 0.0141414141414141
8.9294184 0.0222222222222222
9.8861418 0.0727272727272727
10.8428652 0.15959595959596
11.7995886 0.272727272727273
12.756312 0.387878787878788
13.7130354 0.468686868686869
14.6697588 0.545454545454546
15.6264822 0.614141414141414
16.5832056 0.684848484848485
17.539929 0.74949494949495
18.4966524 0.81010101010101
19.4533758 0.870707070707071
20.4100992 0.898989898989899
21.3668226 0.923232323232323
22.323546 0.943434343434344
23.2802694 0.963636363636364
24.2369928 0.973737373737374
25.1937162 0.98989898989899
26.1504396 0.991919191919192
27.107163 0.997979797979798
28.0638864 0.997979797979798
29.0206098 0.997979797979798
29.9773332 0.997979797979798
30.9340566 0.997979797979798
31.89078 1
};
\addplot [thick, blue, dotted, mark=square*, mark size=1.6pt, mark options={solid}]
table {%
0.3189078 0.0101010101010101
1.2756312 0.0101010101010101
2.2323546 0.0101010101010101
3.189078 0.0101010101010101
4.1458014 0.0101010101010101
5.1025248 0.0101010101010101
6.0592482 0.0101010101010101
7.0159716 0.0101010101010101
7.972695 0.0101010101010101
8.9294184 0.0121212121212121
9.8861418 0.0222222222222222
10.8428652 0.0606060606060606
11.7995886 0.143434343434343
12.756312 0.234343434343434
13.7130354 0.325252525252525
14.6697588 0.466666666666667
15.6264822 0.61010101010101
16.5832056 0.705050505050505
17.539929 0.781818181818182
18.4966524 0.854545454545455
19.4533758 0.909090909090909
20.4100992 0.923232323232323
21.3668226 0.94949494949495
22.323546 0.963636363636364
23.2802694 0.977777777777778
24.2369928 0.97979797979798
25.1937162 0.98989898989899
26.1504396 0.991919191919192
27.107163 0.997979797979798
28.0638864 0.997979797979798
29.0206098 0.997979797979798
29.9773332 0.997979797979798
30.9340566 0.997979797979798
31.89078 1
};
\addplot [thick, color0, dotted, mark=triangle*, mark size=1.6pt, mark options={solid,rotate=180}]
table {%
0.3250418 0.0101010101010101
4.2255434 0.0101010101010101
10.0762958 0.0101010101010101
13.001672 0.0222222222222222
14.9519228 0.0747474747474747
16.9021736 0.18989898989899
18.8524244 0.381818181818182
20.8026752 0.557575757575758
22.752926 0.703030303030303
24.7031768 0.83030303030303
26.6534276 0.929292929292929
28.6036784 0.975757575757576
30.5539292 0.98989898989899
32.50418 1
};

\addplot [thick, color1, dotted, mark=triangle*, mark size=1.6pt, mark options={solid}]
table {%
0.3250418 0.0101010101010101
6.1757942 0.0101010101010101
8.126045 0.0161616161616162
9.1011704 0.0323232323232323
10.0762958 0.0565656565656566
11.0514212 0.0848484848484848
13.001672 0.137373737373737
14.9519228 0.236363636363636
16.9021736 0.339393939393939
18.8524244 0.505050505050505
20.8026752 0.656565656565657
22.752926 0.76969696969697
24.7031768 0.880808080808081
26.6534276 0.955555555555556
28.6036784 0.983838383838384
30.5539292 0.991919191919192
32.50418 1
};
\addplot [thick, red, dotted, mark=diamond*, mark size=1.6pt, mark options={solid}]
table {%
0.3112092 0.0101010101010101
4.9793472 0.0101010101010101
6.8466024 0.0181818181818182
7.78023 0.0404040404040404
8.7138576 0.0606060606060606
9.6474852 0.0868686868686868
10.5811128 0.123232323232323
11.5147404 0.15959595959596
12.448368 0.216161616161616
13.3819956 0.282828282828283
14.3156232 0.371717171717172
15.2492508 0.468686868686869
16.1828784 0.557575757575758
17.116506 0.614141414141414
18.0501336 0.676767676767677
18.9837612 0.729292929292929
19.9173888 0.777777777777778
20.8510164 0.826262626262626
21.784644 0.866666666666666
22.7182716 0.898989898989899
23.6518992 0.939393939393939
24.5855268 0.953535353535353
25.5191544 0.971717171717172
26.452782 0.981818181818182
27.3864096 0.991919191919192
28.3200372 0.997979797979798
29.2536648 0.997979797979798
30.1872924 0.997979797979798
31.12092 1
};

\addplot [thick, black, dotted, mark=triangle*, mark size=1.6pt, mark options={solid,rotate=270}]
table {%
0.2287201 0.0101010101010101
0.9148804 0.0101010101010101
1.6010407 0.0101010101010101
2.287201 0.0101010101010101
2.9733613 0.0101010101010101
3.6595216 0.0101010101010101
4.3456819 0.0181818181818182
5.0318422 0.0323232323232323
5.7180025 0.0444444444444444
6.4041628 0.107070707070707
7.7764834 0.264646464646465
9.148804 0.474747474747475
10.5211246 0.648484848484848
11.8934452 0.787878787878788
13.2657658 0.878787878787878
14.6380864 0.933333333333333
16.010407 0.961616161616161
17.3827276 0.975757575757576
18.7550482 0.98989898989899
20.1273688 0.995959595959596
21.4996894 0.997979797979798
22.87201 1
};
\addplot [thick, color2, dotted, mark=triangle*, mark size=1.6pt, mark options={solid,rotate=90}]
table {%
0.2657687 0.0101010101010101
1.0630748 0.0101010101010101
1.8603809 0.0101010101010101
2.657687 0.0101010101010101
3.4549931 0.0101010101010101
4.2522992 0.0101010101010101
5.0496053 0.0141414141414141
5.8469114 0.0161616161616162
6.6442175 0.0161616161616162
7.4415236 0.0181818181818182
8.2388297 0.0242424242424242
9.0361358 0.0565656565656566
9.8334419 0.098989898989899
10.630748 0.147474747474747
11.4280541 0.18989898989899
12.2253602 0.23030303030303
13.0226663 0.301010101010101
13.8199724 0.37979797979798
14.6172785 0.464646464646465
15.4145846 0.539393939393939
16.2118907 0.656565656565656
17.0091968 0.739393939393939
17.8065029 0.797979797979798
18.603809 0.852525252525252
19.4011151 0.888888888888889
20.1984212 0.931313131313131
20.9957273 0.959595959595959
21.7930334 0.971717171717172
22.5903395 0.981818181818182
23.3876456 0.98989898989899
24.1849517 0.995959595959596
24.9822578 0.995959595959596
25.7795639 0.995959595959596
26.57687 1
};
\addplot [thick, color3, dotted, mark=+, mark size=1.6pt, mark options={solid}]
table {%
0.2574011 0.0101010101010101
4.8906209 0.0161616161616162
5.6628242 0.0464646464646465
6.4350275 0.107070707070707
7.2072308 0.15959595959596
7.9794341 0.238383838383838
8.7516374 0.345454545454545
10.296044 0.511111111111111
11.8404506 0.664646464646464
13.3848572 0.743434343434343
14.9292638 0.826262626262626
16.4736704 0.896969696969697
18.018077 0.933333333333333
19.5624836 0.95959595959596
21.1068902 0.975757575757576
22.6512968 0.995959595959596
24.1957034 0.997979797979798
25.74011 1
};
\addplot [thick, color4, dotted, mark=x, mark size=1.6pt, mark options={solid}]
table {%
0.3506781 0.0114503816793893
1.4027124 0.0114503816793893
2.4547467 0.0114503816793893
3.506781 0.0114503816793893
4.5588153 0.0114503816793893
5.6108496 0.0133587786259542
6.6628839 0.0381679389312977
7.7149182 0.0744274809160305
8.7669525 0.139312977099237
9.8189868 0.248091603053435
10.8710211 0.333969465648855
11.9230554 0.41793893129771
12.9750897 0.522900763358779
14.027124 0.606870229007634
15.0791583 0.67175572519084
16.1311926 0.736641221374046
17.1832269 0.788167938931298
18.2352612 0.82824427480916
19.2872955 0.874045801526718
20.3393298 0.908396946564885
21.3913641 0.935114503816794
22.4433984 0.958015267175572
23.4954327 0.973282442748091
24.547467 0.984732824427481
25.5995013 0.990458015267175
26.6515356 0.990458015267175
27.7035699 0.994274809160305
28.7556042 0.994274809160305
29.8076385 0.99618320610687
30.8596728 0.99618320610687
31.9117071 0.998091603053435
32.9637414 0.998091603053435
34.0157757 0.998091603053435
35.06781 1
};
\end{axis}

\end{tikzpicture}
      	\setlength\belowcaptionskip{0.1cm}
 		\setlength\abovecaptionskip{-0.3cm}
		\caption{\gls{cdf} of the throughput.}
		\label{fig:throughput}
	\end{subfigure}
        \hspace{1.4cm}
	\hfill%
		\begin{subfigure}[t]{0.45\textwidth}
            \hspace{-2.8cm}
		\centering
		\setlength\fwidth{.8\textwidth}
		\setlength\fheight{.45\textwidth}
\begin{tikzpicture}
\pgfplotsset{every tick label/.append style={font=\scriptsize}}

\definecolor{color0}{rgb}{1,0.647058823529412,0}
\definecolor{color1}{rgb}{0.501960784313725,0,0.501960784313725}
\definecolor{color2}{rgb}{0,1,1}
\definecolor{color3}{rgb}{1,0,1}
\definecolor{color4}{rgb}{1,0.752941176470588,0.796078431372549}

\begin{axis}[
width=0.95\fwidth,
height=0.7\fheight,
at={(0\fwidth,0\fheight)},
scale only axis,
tick align=outside,
tick pos=left,
x grid style={white!86.6666666666667!black},
xlabel={Energy Consumption [W]},
xlabel style={font=\footnotesize\color{white!15!black}},
xmajorgrids,
xmin=4000, xmax=8413,
xtick style={color=black},
y grid style={white!86.6666666666667!black},
ylabel={CDF},
ylabel style={font=\footnotesize\color{white!15!black}},
axis background/.style={fill=white},
ymajorgrids,
ymin=0,
ymax=1,
ytick style={color=black},
ytick={-0.2,0,0.2,0.4,0.6,0.8,1,1.2},
yticklabels={−0.2,0.0,0.2,0.4,0.6,0.8,1.0,1.2},
legend style={font=\scriptsize,at={(2.57,1.12)},anchor=south,legend cell align=left,align=left,draw=white!15!black},
legend columns=9,
xlabel shift={-2pt}
]
\addplot [thick, green!50.1960784313725!black, dotted, mark=*, mark size=1.6pt, mark options={solid}]
table {%
81.72 0.0101010101010101
326.88 0.0101010101010101
572.04 0.0101010101010101
817.2 0.0101010101010101
1062.36 0.0101010101010101
1307.52 0.0101010101010101
1552.68 0.0101010101010101
1797.84 0.0101010101010101
2043 0.0101010101010101
2288.16 0.0101010101010101
2533.32 0.0101010101010101
2778.48 0.0101010101010101
3023.64 0.0101010101010101
3268.8 0.0101010101010101
3513.96 0.0101010101010101
3759.12 0.0101010101010101
4004.28 0.0101010101010101
4494.6 0.0101010101010101
4739.76 0.0202020202020202
4984.92 0.0363636363636364
5230.08 0.0545454545454545
5475.24 0.0767676767676768
5720.4 0.165656565656566
5965.56 0.248484848484849
6210.72 0.363636363636364
6455.88 0.515151515151515
6701.04 0.612121212121212
6946.2 0.729292929292929
7191.36 0.852525252525252
7436.52 0.943434343434343
7681.68 0.975757575757576
7926.84 0.987878787878788
8172 1
};
\addplot [thick, blue, dotted, mark=square*, mark size=1.6pt, mark options={solid}]
table {%
82.46 0.0101010101010101
329.84 0.0101010101010101
577.22 0.0101010101010101
824.6 0.0101010101010101
1071.98 0.0101010101010101
1319.36 0.0101010101010101
1566.74 0.0101010101010101
1814.12 0.0101010101010101
2061.5 0.0101010101010101
2308.88 0.0101010101010101
2556.26 0.0101010101010101
2803.64 0.0101010101010101
3051.02 0.0101010101010101
3298.4 0.0101010101010101
3545.78 0.0101010101010101
3793.16 0.0101010101010101
4040.54 0.0101010101010101
4535.3 0.0101010101010101
5030.06 0.0101010101010101
5277.44 0.0181818181818182
5524.82 0.0303030303030303
5772.2 0.0787878787878788
6019.58 0.141414141414141
6266.96 0.246464646464646
6514.34 0.391919191919192
6761.72 0.541414141414141
7009.1 0.682828282828283
7256.48 0.828282828282828
7503.86 0.927272727272727
7751.24 0.973737373737374
7998.62 0.993939393939394
8246 1
};
\addplot [thick, color0, dotted, mark=triangle*, mark size=1.6pt, mark options={solid,rotate=180}]
table {%
83.09 0.0101010101010101
332.36 0.0101010101010101
581.63 0.0101010101010101
830.9 0.0101010101010101
1080.17 0.0101010101010101
1329.44 0.0101010101010101
1578.71 0.0101010101010101
1827.98 0.0101010101010101
2077.25 0.0101010101010101
2326.52 0.0101010101010101
2575.79 0.0101010101010101
2825.06 0.0101010101010101
3074.33 0.0101010101010101
3323.6 0.0101010101010101
3572.87 0.0101010101010101
3822.14 0.0101010101010101
4071.41 0.0101010101010101
4320.68 0.0101010101010101
4819.22 0.0121212121212121
5317.76 0.0262626262626263
5567.03 0.0363636363636364
5816.3 0.0666666666666667
6065.57 0.137373737373737
6314.84 0.244444444444444
6564.11 0.367676767676768
6813.38 0.529292929292929
7062.65 0.707070707070707
7311.92 0.836363636363636
7561.19 0.933333333333333
7810.46 0.97979797979798
8059.73 0.997979797979798
8309 1
};
\addplot [thick, red, dotted, mark=diamond*, mark size=1.6pt, mark options={solid}]
table {%
79.9 0.0101010101010101
319.6 0.0101010101010101
559.3 0.0101010101010101
799 0.0101010101010101
1038.7 0.0101010101010101
1278.4 0.0101010101010101
1518.1 0.0101010101010101
1757.8 0.0101010101010101
1997.5 0.0101010101010101
2237.2 0.0101010101010101
2476.9 0.0101010101010101
2716.6 0.0101010101010101
2956.3 0.0101010101010101
3196 0.0101010101010101
3435.7 0.0121212121212121
3675.4 0.0121212121212121
3915.1 0.0121212121212121
4154.8 0.0121212121212121
4394.5 0.0121212121212121
4634.2 0.0181818181818182
4873.9 0.0343434343434343
5113.6 0.0727272727272727
5353.3 0.111111111111111
5593 0.15959595959596
5832.7 0.238383838383838
6072.4 0.335353535353535
6312.1 0.450505050505051
6551.8 0.571717171717172
6791.5 0.705050505050505
7031.2 0.828282828282828
7270.9 0.884848484848485
7510.6 0.935353535353535
7750.3 0.977777777777778
7990 1
};
\addplot [thick, color1, dotted, mark=triangle*, mark size=1.6pt, mark options={solid}]
table {%
83.09 0.0101010101010101
332.36 0.0101010101010101
581.63 0.0101010101010101
830.9 0.0101010101010101
1080.17 0.0101010101010101
1329.44 0.0101010101010101
1578.71 0.0101010101010101
1827.98 0.0101010101010101
2077.25 0.0101010101010101
2326.52 0.0101010101010101
2575.79 0.0101010101010101
2825.06 0.0101010101010101
3074.33 0.0101010101010101
3323.6 0.0101010101010101
3572.87 0.0101010101010101
3822.14 0.0101010101010101
4071.41 0.0101010101010101
4569.95 0.0101010101010101
5068.49 0.0202020202020202
5317.76 0.0363636363636364
5567.03 0.0606060606060606
5816.3 0.101010101010101
6065.57 0.193939393939394
6314.84 0.331313131313131
6564.11 0.446464646464647
6813.38 0.581818181818182
7062.65 0.743434343434343
7311.92 0.856565656565656
7561.19 0.937373737373737
7810.46 0.97979797979798
8059.73 0.997979797979798
8309 1
};
\addplot [thick, black, dotted, mark=triangle*, mark size=1.6pt, mark options={solid,rotate=270}]
table {%
74.92 0.0101010101010101
299.68 0.0101010101010101
524.44 0.0101010101010101
749.2 0.0101010101010101
973.96 0.0101010101010101
1198.72 0.0101010101010101
1423.48 0.0101010101010101
1648.24 0.0101010101010101
1873 0.0101010101010101
2097.76 0.0101010101010101
2322.52 0.0101010101010101
2547.28 0.0121212121212121
2772.04 0.0121212121212121
2996.8 0.0181818181818182
3221.56 0.0262626262626263
3446.32 0.0323232323232323
3671.08 0.0464646464646465
3895.84 0.0868686868686869
4120.6 0.137373737373737
4345.36 0.17979797979798
4570.12 0.272727272727273
4794.88 0.365656565656566
5019.64 0.442424242424242
5244.4 0.557575757575758
5469.16 0.664646464646465
5693.92 0.75959595959596
5918.68 0.84040404040404
6143.44 0.901010101010101
6368.2 0.929292929292929
6592.96 0.967676767676768
6817.72 0.983838383838384
7042.48 0.991919191919192
7267.24 0.995959595959596
7492 1
};
\addplot [thick, color2, dotted, mark=triangle*, mark size=1.6pt, mark options={solid,rotate=90}]
table {%
84.13 0.0101010101010101
336.52 0.0101010101010101
588.91 0.0101010101010101
841.3 0.0101010101010101
1093.69 0.0101010101010101
1346.08 0.0101010101010101
1598.47 0.0101010101010101
1850.86 0.0101010101010101
2103.25 0.0101010101010101
2355.64 0.0101010101010101
2608.03 0.0101010101010101
2860.42 0.0101010101010101
3112.81 0.0101010101010101
3365.2 0.0101010101010101
3617.59 0.0101010101010101
3869.98 0.0101010101010101
4122.37 0.0121212121212121
4374.76 0.0121212121212121
4627.15 0.0181818181818182
4879.54 0.0404040404040404
5131.93 0.0707070707070707
5384.32 0.113131313131313
5636.71 0.16969696969697
5889.1 0.248484848484848
6141.49 0.357575757575758
6393.88 0.458585858585859
6646.27 0.597979797979798
6898.66 0.703030303030303
7151.05 0.822222222222222
7403.44 0.898989898989899
7655.83 0.951515151515151
7908.22 0.981818181818182
8160.61 0.995959595959596
8413 1
};
\addplot [thick, color3, dotted, mark=+, mark size=1.6pt, mark options={solid}]
table {%
75.96 0.0101010101010101
303.84 0.0101010101010101
531.72 0.0101010101010101
759.6 0.0101010101010101
987.48 0.0101010101010101
1215.36 0.0101010101010101
1443.24 0.0101010101010101
1671.12 0.0101010101010101
1899 0.0101010101010101
2126.88 0.0101010101010101
2354.76 0.0101010101010101
2582.64 0.0101010101010101
2810.52 0.0121212121212121
3038.4 0.0121212121212121
3266.28 0.0161616161616162
3494.16 0.0242424242424242
3722.04 0.0323232323232323
3949.92 0.0505050505050505
4177.8 0.0747474747474747
4405.68 0.105050505050505
4633.56 0.157575757575758
4861.44 0.232323232323232
5089.32 0.333333333333333
5317.2 0.456565656565657
5545.08 0.563636363636364
5772.96 0.654545454545455
6000.84 0.739393939393939
6228.72 0.828282828282828
6456.6 0.898989898989899
6684.48 0.941414141414141
6912.36 0.967676767676768
7140.24 0.983838383838384
7368.12 0.98989898989899
7596 1
};
\addplot [thick, color4, dotted, mark=x, mark size=1.6pt, mark options={solid}]
table {%
80.21 0.0114503816793893
320.84 0.0114503816793893
561.47 0.0114503816793893
802.1 0.0114503816793893
1042.73 0.0114503816793893
1283.36 0.0114503816793893
1523.99 0.0114503816793893
1764.62 0.0114503816793893
2005.25 0.0114503816793893
2245.88 0.0114503816793893
2486.51 0.0114503816793893
2727.14 0.0114503816793893
2967.77 0.0114503816793893
3208.4 0.0133587786259542
3449.03 0.0152671755725191
3689.66 0.017175572519084
3930.29 0.0267175572519084
4170.92 0.0419847328244275
4411.55 0.066793893129771
4652.18 0.0935114503816794
4892.81 0.129770992366412
5133.44 0.200381679389313
5374.07 0.297709923664122
5614.7 0.379770992366412
5855.33 0.49618320610687
6095.96 0.608778625954198
6336.59 0.738549618320611
6577.22 0.818702290076336
6817.85 0.883587786259542
7058.48 0.948473282442748
7299.11 0.973282442748092
7539.74 0.986641221374046
7780.37 0.998091603053435
8021 1
};
\end{axis}

\end{tikzpicture}
    	\setlength\belowcaptionskip{0.1cm}
 		\setlength\abovecaptionskip{0.1cm}
            \caption{\gls{cdf} of the energy consumption.}
            \label{fig:EnCOnsumption}
	\end{subfigure}%
        \hfill
		\begin{subfigure}[t]{0.45\textwidth}
		\centering
		\setlength\fwidth{.8\textwidth}
		\setlength\fheight{.45\textwidth}
\begin{tikzpicture}
\pgfplotsset{every tick label/.append style={font=\scriptsize}}

\definecolor{color0}{rgb}{1,0.647058823529412,0}
\definecolor{color1}{rgb}{0.501960784313725,0,0.501960784313725}
\definecolor{color2}{rgb}{0,1,1}
\definecolor{color3}{rgb}{1,0,1}
\definecolor{color4}{rgb}{1,0.752941176470588,0.796078431372549}

\begin{axis}[
width=0.95\fwidth,
height=0.7\fheight,
at={(0\fwidth,0\fheight)},
scale only axis,
tick align=outside,
tick pos=left,
x grid style={white!86.6666666666667!black},
xlabel={Radio Link Failure (RLF) [Value]},
xlabel style={font=\footnotesize\color{white!15!black}},
xmajorgrids,
xmin=0, xmax=6,
xtick style={color=black},
y grid style={white!86.6666666666667!black},
ylabel={CDF},
ylabel style={font=\footnotesize\color{white!15!black}},
axis background/.style={fill=white},
ymajorgrids,
ymin=0.2, 
ymax=1,
ytick style={color=black},
ytick={0.0, 0.2,0.4,0.6,0.8,1.0},
yticklabels={0.0, 0.2,0.4,0.6,0.8,1.0},
legend style={font=\scriptsize,at={(1.2,1.12)},anchor=south,legend cell align=left,align=left,draw=white!15!black},
legend columns=9,
xlabel shift={-2pt}
]
\addplot [thick, green!50.1960784313725!black, dotted, mark=*, mark size=1.6pt, mark options={solid}]
table {%
0.03 0.771717171717172
0.21 0.771717171717172
0.39 0.771717171717172
0.57 0.771717171717172
0.75 0.771717171717172
0.93 0.771717171717172
1.11 0.97979797979798
1.29 0.97979797979798
1.65 0.97979797979798
1.83 0.97979797979798
2.01 0.995959595959596
2.19 0.995959595959596
2.37 0.995959595959596
2.55 0.995959595959596
2.73 0.995959595959596
2.91 0.995959595959596
};
\addplot [thick, blue, dotted, mark=square*, mark size=1.6pt, mark options={solid}]
table {%
0.03 0.715151515151515
0.27 0.715151515151515
0.51 0.715151515151515
0.75 0.715151515151515
0.99 0.715151515151515
1.23 0.937373737373737
1.47 0.937373737373737
1.71 0.937373737373737
1.95 0.937373737373737
2.19 0.991919191919192
2.43 0.991919191919192
2.67 0.991919191919192
2.91 0.991919191919192
};
\addplot [thick, color0, dotted, mark=triangle*, mark size=1.6pt, mark options={solid,rotate=180}]
table {%
0.01 0.997979797979798
0.21 0.997979797979798
0.41 0.997979797979798
0.61 0.997979797979798
0.81 0.997979797979798
1.01 0.997979797979798
};
\addplot [thick, red, dotted, mark=diamond*, mark size=1.6pt, mark options={solid}]
table {%
0.02 0.929292929292929
0.34 0.929292929292929
0.66 0.929292929292929
0.98 0.929292929292929
1.14 0.991919191919192
1.46 0.991919191919192
1.78 0.991919191919192
2.04 0.991919191919192
};
\addplot [thick, color1, dotted, mark=triangle*, mark size=1.6pt, mark options={solid}]
table {%
0.06 0.894949494949495
0.42 0.894949494949495
0.78 0.894949494949495
1.14 0.915151515151515
1.5 0.915151515151515
1.86 0.915151515151515
2.22 0.943434343434343
2.58 0.943434343434343
2.94 0.943434343434343
3.3 0.963636363636364
3.66 0.963636363636364
4.02 0.97979797979798
4.38 0.97979797979798
4.74 0.97979797979798
5.1 0.991919191919192
5.46 0.991919191919192
5.82 0.991919191919192
};
\addplot [thick, black, dotted, mark=triangle*, mark size=1.6pt, mark options={solid,rotate=270}]
table {%
0.08 0.270707070707071
0.56 0.270707070707071
1.04 0.591919191919192
1.52 0.591919191919192
2 0.591919191919192
2.48 0.804040404040404
2.96 0.804040404040404
3.44 0.913131313131313
3.92 0.913131313131313
4.4 0.971717171717172
4.88 0.971717171717172
5.36 0.991919191919192
5.84 0.991919191919192
6.32 0.995959595959596
6.8 0.995959595959596
7.28 0.997979797979798
7.76 0.997979797979798
};
\addplot [thick, color2, dotted, mark=triangle*, mark size=1.6pt, mark options={solid,rotate=90}]
table {%
0.06 0.777777777777778
0.42 0.777777777777778
0.78 0.777777777777778
1.14 0.907070707070707
1.5 0.907070707070707
1.86 0.907070707070707
2.22 0.94949494949495
2.58 0.94949494949495
2.94 0.94949494949495
3.3 0.977777777777778
3.66 0.977777777777778
4.02 0.993939393939394
4.38 0.993939393939394
4.74 0.993939393939394
5.1 0.997979797979798
5.46 0.997979797979798
5.82 0.997979797979798
};
\addplot [thick, color3, dotted, mark=+, mark size=1.6pt, mark options={solid}]
table {%
0.04 0.64040404040404
0.28 0.64040404040404
0.52 0.64040404040404
0.76 0.64040404040404
1 0.64040404040404
1.24 0.888888888888889
1.48 0.888888888888889
1.72 0.888888888888889
1.96 0.888888888888889
2.2 0.967676767676768
2.44 0.967676767676768
2.68 0.967676767676768
2.92 0.967676767676768
3.16 0.991919191919192
3.4 0.991919191919192
3.64 0.991919191919192
3.88 0.991919191919192
};
\addplot [thick, color4, dotted, mark=x, mark size=1.6pt, mark options={solid}]
table {%
0.04 0.761450381679389
0.28 0.761450381679389
0.52 0.761450381679389
0.76 0.761450381679389
1 0.761450381679389
1.24 0.919847328244275
1.48 0.919847328244275
1.72 0.919847328244275
1.96 0.919847328244275
2.2 0.975190839694656
2.44 0.975190839694656
2.68 0.975190839694656
2.92 0.975190839694656
3.16 0.990458015267175
3.4 0.990458015267175
3.64 0.990458015267175
3.88 0.990458015267175
};
\end{axis}

\end{tikzpicture}
    	\setlength\belowcaptionskip{0.1cm}
 		\setlength\abovecaptionskip{0.1cm}
            \caption{\gls{cdf} of the \gls{rlf}.}
            \label{fig:RLF}
	\end{subfigure}%
 	\hfill%
	\begin{subfigure}[t]{0.45\textwidth}
            \hspace{-0.8cm}
            \centering
		\setlength\fwidth{.8\textwidth}
		\setlength\fheight{.45\textwidth}
\begin{tikzpicture}
\pgfplotsset{every tick label/.append style={font=\scriptsize}}

\definecolor{color0}{rgb}{1,0.647058823529412,0}
\definecolor{color1}{rgb}{0.501960784313725,0,0.501960784313725}
\definecolor{color2}{rgb}{0,1,1}
\definecolor{color3}{rgb}{1,0,1}
\definecolor{color4}{rgb}{1,0.752941176470588,0.796078431372549}

\begin{axis}[
width=0.95\fwidth,
height=0.7\fheight,
at={(0\fwidth,0\fheight)},
scale only axis,
tick align=outside,
tick pos=left,
x grid style={white!86.6666666666667!black},
xlabel={Cost of activating a cell $\delta$},
xlabel style={font=\footnotesize\color{white!15!black}},
xmajorgrids,
xmin=0, xmax=2.7,
xtick style={color=black},
y grid style={white!86.6666666666667!black},
ylabel={CDF},
ylabel style={font=\footnotesize\color{white!15!black}},
axis background/.style={fill=white},
ymajorgrids,
ymin=0, 
ymax=1,
ytick style={color=black},
ytick={-0.2,0,0.2,0.4,0.6,0.8,1,1.2},
yticklabels={−0.2,0.0,0.2,0.4,0.6,0.8,1.0,1.2},
legend style={font=\scriptsize,at={(2.57,1.12)},anchor=south,legend cell align=left,align=left,draw=white!15!black},
legend columns=9,
xlabel shift={-2pt}
]
\addplot [thick, green!50.1960784313725!black, dotted, mark=*, mark size=1.6pt, mark options={solid}]
table {%
0.008 0.523232323232323
0.108 0.668686868686869
0.208 0.783838383838384
0.308 0.816161616161616
0.408 0.844444444444444
0.508 0.868686868686868
0.608 0.886868686868687
0.708 0.901010101010101
0.808 0.915151515151515
0.908 0.929292929292929
1.008 0.939393939393939
1.108 0.959595959595959
1.208 0.959595959595959
1.308 0.969696969696969
1.408 0.979797979797979
1.508 0.979797979797979
1.608 0.989898989898989
1.708 0.989898989898989
};
\addplot [thick, blue, dotted, mark=square*, mark size=1.6pt, mark options={solid}]
table {%
0.008 0.131313131313131
0.108 0.391919191919192
0.208 0.501010101010101
0.308 0.597979797979798
0.408 0.668686868686869
0.508 0.731313131313131
0.608 0.787878787878788
0.708 0.838383838383838
0.808 0.878787878787878
0.908 0.919191919191919
1.008 0.935353535353535
1.108 0.959595959595959
1.208 0.959595959595959
1.308 0.969696969696969
1.408 0.979797979797979
1.508 0.979797979797979
1.608 0.989898989898989
1.708 0.989898989898989
};
\addplot [thick, color0, dotted, mark=triangle*, mark size=1.6pt, mark options={solid,rotate=180}]
table {%
0.01 1
0.11 1
0.21 1
0.31 1
0.41 1
0.51 1
};
\addplot [thick, red, dotted, mark=diamond*, mark size=1.6pt, mark options={solid}]
table {%
0.027 0.0767676767676768
0.162 0.0848484848484848
0.297 0.127272727272727
0.432 0.17979797979798
0.567 0.284848484848485
0.702 0.353535353535353
0.837 0.58989898989899
0.972 0.862626262626263
1.107 0.862626262626263
1.242 0.862626262626263
1.377 0.862626262626263
1.512 0.862626262626263
1.647 0.864646464646465
1.782 0.866666666666667
1.917 0.868686868686869
2.052 0.870707070707071
2.187 0.872727272727273
2.322 0.874747474747475
2.457 0.886868686868687
2.592 0.907070707070707
};
\addplot [thick, color1, dotted, mark=triangle*, mark size=1.6pt, mark options={solid}]
table {%
0.027 0.834343434343434
0.162 0.876767676767677
0.297 0.896969696969697
0.432 0.911111111111111
0.567 0.921212121212121
0.702 0.933333333333334
0.837 0.943434343434344
0.972 0.951515151515152
1.107 0.957575757575758
1.242 0.963636363636364
1.377 0.96969696969697
1.512 0.975757575757576
1.647 0.983838383838385
1.782 0.989898989898991
1.917 0.991919191919193
2.052 0.993939393939395
2.187 0.995959595959597
2.322 0.995959595959597
2.457 0.997979797979799
2.592 0.997979797979799
};
\addplot [thick, black, dotted, mark=triangle*, mark size=1.6pt, mark options={solid,rotate=270}]
table {%
0.027 0.0202020202020202
0.162 0.0202020202020202
0.297 0.0202020202020202
0.432 0.0202020202020202
0.567 0.0202020202020202
0.702 0.0202020202020202
0.837 0.0222222222222222
0.972 0.0242424242424242
1.107 0.0242424242424242
1.242 0.0262626262626263
1.377 0.0282828282828283
1.512 0.0363636363636364
1.647 0.0525252525252525
1.782 0.0969696969696969
1.917 0.165656565656566
2.052 0.246464646464646
2.187 0.345454545454545
2.322 0.462626262626263
2.457 0.622222222222222
2.592 0.947474747474747
};
\addplot [thick, color2, dotted, mark=triangle*, mark size=1.6pt, mark options={solid,rotate=90}]
table {%
0.027 0.0909090909090909
0.162 0.103030303030303
0.297 0.117171717171717
0.432 0.151515151515152
0.567 0.24040404040404
0.702 0.298989898989899
0.837 0.563636363636364
0.972 0.826262626262626
1.107 0.826262626262626
1.242 0.826262626262626
1.377 0.826262626262626
1.512 0.826262626262626
1.647 0.826262626262626
1.782 0.826262626262626
1.917 0.826262626262626
2.052 0.826262626262626
2.187 0.83030303030303
2.322 0.832323232323232
2.457 0.846464646464646
2.592 0.872727272727273
};
\addplot [thick, color3, dotted, mark=+, mark size=1.6pt, mark options={solid}]
table {%
0.027 0.101010101010101
0.162 0.101010101010101
0.297 0.101010101010101
0.432 0.101010101010101
0.567 0.103030303030303
0.702 0.103030303030303
0.837 0.109090909090909
0.972 0.131313131313131
1.107 0.131313131313131
1.242 0.135353535353535
1.377 0.143434343434343
1.512 0.155555555555556
1.647 0.17979797979798
1.782 0.224242424242424
1.917 0.23030303030303
2.052 0.252525252525253
2.187 0.284848484848485
2.322 0.343434343434344
2.457 0.519191919191919
2.592 0.808080808080808
};
\addplot [thick, color4, dotted, mark=x, mark size=1.6pt, mark options={solid}]
table {%
0.027 0.125954198473282
0.162 0.125954198473282
0.297 0.125954198473282
0.432 0.131679389312977
0.567 0.156488549618321
0.702 0.166030534351145
0.837 0.305343511450382
0.972 0.553435114503817
1.107 0.553435114503817
1.242 0.553435114503817
1.377 0.553435114503817
1.512 0.561068702290076
1.647 0.57442748091603
1.782 0.618320610687023
1.917 0.618320610687023
2.052 0.622137404580153
2.187 0.641221374045801
2.322 0.669847328244275
2.457 0.738549618320611
2.592 0.854961832061069
};
\end{axis}

\end{tikzpicture}
    	\setlength\belowcaptionskip{0.1cm}
 		\setlength\abovecaptionskip{0.1cm}
            \caption{\gls{cdf} of the cost associated to activating a cell.}
            \label{fig:CostCellON}
	\end{subfigure}
		\hfill%
        \setlength\belowcaptionskip{-.3cm}
        \caption{Comparison across multiple \glspl{kpm} for the baselines, PPO, and DQN agents.}
        \label{fig:EEKPICOmparison}
\end{figure*}

We also implemented four different baseline strategies to activate and deactivate the \glspl{bs} and generate a training dataset for the \gls{drl} agents. 
\textbf{Random Policy} deactivates a random number of \glspl{bs} every second, following a uniform random distribution.
\textbf{Static and Dynamic Policies} are based on heuristics discussed in~\cite{7060678}. Both algorithms determine the numbers of \glspl{bs} to activate and deactivate every E2 indication periodicity (i.e., 100\,ms). The heuristics are based on parameters $N_{\rm On, P1}$, $N_{\rm On, P2}$, $N_{\rm Off}$, whose sum is $N$, the total number of NR \glspl{bs} in the scenario, and discussed below. 
The algorithms calculate the number of users within the coverage area of each \gls{bs}, taking into account a predefined \gls{sinr} target (always equal to 13\,dB) and a radius $R = 2000$\,m.
\glspl{bs} are ranked based on the highest amount of connected \glspl{ue} with \gls{sinr} higher than the targeted value. 
%
%
The algorithms then activate the first $N_{\rm On,P1}$ \glspl{bs} and do not consider them in the following steps.
The static policy then sorts the remaining \glspl{bs} based on the shortest distance between them and the closest \gls{ue}. 
The dynamic policy, instead, ranks the \glspl{bs} based on the time it would take for a \gls{ue} to reach the location of the \gls{bs}, thus also including the speed of the \gls{ue} in the evaluation. 
For both policies, the first $N_{\rm On,P2}$ cells are activated. Finally, the remaining $N_{\rm Off}$ cells are deactivated.
extbf{Always On} keeps all \glspl{bs} active throughout the entire simulation. This represents a baseline where both the coverage and energy consumption are at their maximum levels.
\subsection{Training Dataset}

The dataset is composed of approximately 3000\, simulations using the heuristics discussed above, each simulation is 10\,s long and metrics data were saved every 100\,ms. 
Each heuristic variant has been associated with both uniform and non-uniform positioning of \glspl{ue} for training, while in testing we consider only uniformly allocated \glspl{ue}. 
Finally, we include different combinations of the $(N_{\rm On, P1}, N_{\rm On, P2}, N_{\rm Off})$ tuple for the static and dynamic heuristics, i.e., $(4,2,1), (3,2,2), (2,2,3)$. 


\begin{figure}
    \centering
    \resizebox{0.46\textwidth}{!}{
\begin{tikzpicture}
\pgfplotsset{every tick label/.append style={font=\scriptsize}}

\definecolor{color0}{rgb}{1,0.647058823529412,0}
\definecolor{color1}{rgb}{0.501960784313725,0,0.501960784313725}
\definecolor{color2}{rgb}{0,1,1}
\definecolor{color3}{rgb}{1,0,1}
\definecolor{color4}{rgb}{1,0.752941176470588,0.796078431372549}

\begin{axis}[
width=8cm,
height=5cm,
tick align=outside,
tick pos=left,
x grid style={white!50.1960784313725!black},
xlabel style={font=\footnotesize\color{white!15!black}},
ylabel style={font=\footnotesize\color{white!15!black}},
xlabel={Average Throughput [\%]},
xmajorgrids,
xmin=45.1618136007987, xmax=100,
xminorgrids,
xtick style={color=black},
y grid style={white!50.1960784313725!black},
ylabel={\parbox{4cm}{\centering Average Energy\\Consumption [\%]}},
ymajorgrids,
ymin=74.6246760400198, ymax=100,
yminorgrids,
ytick style={color=black},
xlabel shift={-2pt}
]
\addplot [semithick, green!50.1960784313725!black, mark=*, mark size=3, mark options={solid}, only marks]
table {%
72.4089967534879 95.7503297923202
};
\addplot [semithick, blue, mark=square*, mark size=3, mark options={solid}, only marks]
table {%
74.5212583308336 99.3430209764277
};
\addplot [semithick, color0, mark=triangle*, mark size=3, mark options={solid,rotate=180}, only marks]
table {%
100 100
};
\addplot [semithick, red, mark=diamond*, mark size=3, mark options={solid}, only marks]
table {%
78.6752312782215 94.5926481611314
};
\addplot [semithick, color1, mark=triangle*, mark size=3, mark options={solid}, only marks]
table {%
91.3560181806838 98.4843793269146
};
\addplot [semithick, black, mark=triangle*, mark size=3, mark options={solid,rotate=270}, only marks]
table {%
47.7731558102845 75.8330248000189
};
\addplot [semithick, color2, mark=triangle*, mark size=3, mark options={solid,rotate=90}, only marks]
table {%
72.5403169945124 95.6040213757536
};
\addplot [semithick, color3, mark=+, mark size=3, mark options={solid}, only marks]
table {%
53.8211396483012 80.769790114532
};
\addplot [semithick, color4, mark=x, mark size=3, mark options={solid}, only marks]
table {%
66.0064798849747 86.4139952668066
};
\draw (axis cs:72.4089967534879,91.7503297923202) node[
  scale=0.9,
  anchor=south,
  text=green!50.1960784313725!black,
  rotate=0.0
]{Static};
\draw (axis cs:68.5212583308336,96.8430209764277) node[
  scale=0.9,
  anchor=south,
  text=blue,
  rotate=0.0
]{Dynamic};
\draw (axis cs:99.7,94) node[
  scale=0.9,
  anchor=south,
  text=color0,
  rotate=90.0
]{Always On};
\draw (axis cs:79.6,90.5926481611314) node[
  scale=0.9,
  anchor=south,
  text=red,
  rotate=0.0
]{PPO-1};
\draw (axis cs:91.3560181806838,94.4843793269146) node[
  scale=0.9,
  anchor=south,
  text=color1,
  rotate=0.0
]{DQN};
\draw (axis cs:48.7731558102845,76.8330248000189) node[
  scale=0.9,
  anchor=south,
  text=black,
  rotate=0.0
]{PPO-2};
\draw (axis cs:67.0403169945124,94.6040213757536) node[
  scale=0.9,
  anchor=south,
  text=color2,
  rotate=0.0
]{PPO-3};
\draw (axis cs:53.8211396483012,81.769790114532) node[
  scale=0.9,
  anchor=south,
  text=color3,
  rotate=0.0
]{PPO-4};
\draw (axis cs:66.0064798849747,82.9139952668066) node[
  scale=0.9,
  anchor=south,
  text=color4,
  rotate=0.0
]{PPO-5};
\end{axis}

\end{tikzpicture}}
    \setlength\belowcaptionskip{-.3cm}
    \caption{Percentage of the average Throughput compared to the percentage of the average Energy Consumption.}
    \label{fig:ThrVSEnCOns}
\end{figure}
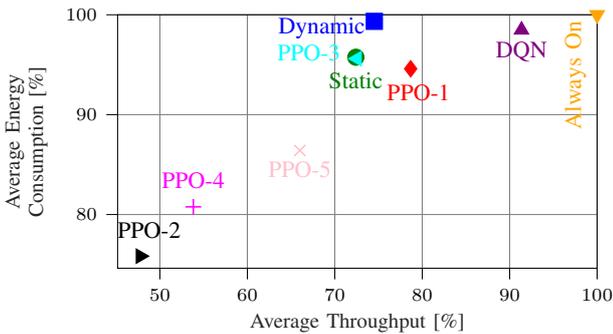

\section{Results}
\label{sec:results}
In this section, we report results on the 6\,\gls{drl} agents that have been configured with the weights in Table~\ref{tab:RFweights} and parameters discussed in Sec.~\ref{sec:drl-agent-design}. 
The results are based on simulation runs which are statistically independent compared to those in the training dataset. 
In these simulations, the \gls{drl} agents perform online inference and control the activation and deactivation of the cell. 
Experiments are repeated with different seeds for a total of 500\,data points for each agent or baseline. 
In Fig.~\ref{fig:EEKPICOmparison}, we plot \glspl{cdf} of the total values for the throughput $\rho$, the energy consumption $\gamma$, the number of \glspl{rlf} per \gls{bs}, and the cost factor $\delta$ associated to cell activation, defined in Sec.~\ref{sec:system}. 
The \glspl{cdf} considers the values for \glspl{kpm} in each E2 report.  
Figure~\ref{fig:ThrVSEnCOns}, instead, considers the average values for the throughput and energy consumption and reports the ratio between both \glspl{kpm} for each configuration and those for the \textit{Always On} policy.
Plots show that the \textit{Always On} policy offers the maximum throughput, energy consumption, and coverage (i.e., no \glspl{rlf} for the \glspl{ue}) without the cost of activating cells, as expected. 
On the other hand, the PPO-2 agent offers the best energy savings (Fig.~\ref{fig:EnCOnsumption}), 24.2\% less than \textit{Always On}.
Consequently, the number of cells deactivated by the model is the highest, resulting in lower throughput (52.3\% lower compared to the \textit{Always On} policy). 
\textit{Static} and \textit{Dynamic} baselines allow the system to save 0.7\% and 4.3\% more energy compared to the \textit{Always On} policy, by decreasing the throughput of 23.5\% and 27.6\%. 
The throughput decrease is associated with a lower coverage with at most 3\,\glspl{ue} in \gls{rlf} for both baselines. 
These baselines are conservative in terms of cell activation and deactivation, thus with a low cost for activating cells. 
DQN and PPO-1 agents offer higher throughput compared to the \textit{Static} and \textit{Dynamic} baselines, while also saving more energy than the \textit{Dynamic} baseline. 
Notably, the PPO-1 agent dominates both baselines, providing higher energy savings (1.2\% and 4.8\% improvement) and higher throughput (7.2\% and 3.1\% respectively). 
Both models have similar coverage to the baselines, with at most 2 and 1 \glspl{ue} in \gls{rlf} for PPO-1 and DQN, respectively.
Regarding the cell activation cost, DQN takes a conservative approach to changing the status for each \gls{bs}, outperforming both \textit{Static} and \textit{Dynamic} baselines.
Finally, PPO-4 and PPO-5 agents provide 46.2\% and 34\% less throughput, respectively, compared to the \textit{Always On} policy but the amount of energy saved is extremely high: the former saves 19.3\% and the latter 13.6\%. 

Overall, Fig.~\ref{fig:ThrVSEnCOns} indicates that if the goal is to save energy, a quantity of throughput has to be sacrificed, particularly in scenarios with limited user traffic variability and limited mobility (i.e., \glspl{ue} rarely leave a cell coverage area during the simulations). 
In this case, it is up to the operator to decide the optimal trade-off based on the specific intent and business requirements.
For instance, the PPO-2 agent can be useful in situations where energy saving is the top priority, while PPO-4 and PPO-5 consume more but provide higher throughput. 
Alternatively, if energy saving is needed without losing too much throughput and coverage, PPO-1 and DQN are the best trade-offs between these two metrics investigated in this paper. 

\section{Conclusions}
\label{sec:conclusions}
This paper introduces a novel approach to address the dual challenge of meeting the escalating data demands of \gls{5g} networks while managing their substantial energy consumption.
We propose an intelligent, programmable solution for energy-efficient Open \gls{ran} networks without compromising service quality.
The approach studied in this paper involves using \gls{drl} algorithms, specifically \gls{ppo} and \gls{dqn}, to dynamically control the activation and deactivation of \gls{5g} cells based on historical data.
Through comprehensive simulation campaigns and comparison with policy heuristics, results demonstrate the effectiveness of the trained models in maximizing a reward function encapsulating throughput, energy consumption, coverage, and cell activation cost.
Notably, results show that these models offer a trade-off between performance and energy efficiency, highlighting the need for intelligent energy-saving mechanisms in future Open \gls{ran} deployments.
By contributing to the development of such mechanisms, the aim is to improve the efficiency and sustainability of \gls{5g} networks in the face of growing data demand and environmental concerns.

\footnotesize
\bibliographystyle{IEEEtran}
\bibliography{biblio}

\end{document}